\documentclass[onecolumn,journal]{IEEEtran}
\IEEEoverridecommandlockouts

\usepackage{amsmath}
\usepackage{amsfonts}
\usepackage{amsthm}   
\usepackage{mathrsfs}
\usepackage{tikz}
\usepackage[utf8]{inputenc}

\usepackage{epstopdf}
\usepackage{mathtools}
\usepackage{graphicx}
\usepackage{bbm}
\usepackage{enumerate}
\usepackage{epsf,verbatim,amssymb,array,cite,multicol,multirow}  
\usepackage{psfrag,bm,xspace}
\usepackage{hhline}
\usepackage{xstring}
\usepackage{ifthen}

\ifdefined \theorem 
\else
  \newtheorem{theorem}{Theorem}
\fi
\newtheorem{lem}{Lemma}
\ifdefined \corollary 
\else
\newtheorem{corollary}{Corollary}
\fi

\ifdefined \proposition 
\else
\newtheorem{proposition}{Proposition}
\fi

\ifdefined \definition 
\else
\newtheorem{definition}{Definition}
\fi

\ifdefined \example 
\else
\newtheorem{example}{Example}
\fi

\ifdefined \remark 
\else
\newtheorem{remark}{Remark}
\fi

\newtheorem{innercustomgeneric}{\customgenericname}
\providecommand{\customgenericname}{}
\newcommand{\newcustomtheorem}[2]{%
  \newenvironment{#1}[1]
  {%
   \renewcommand\customgenericname{#2}%
   \renewcommand\theinnercustomgeneric{##1}%
   \innercustomgeneric
  }
  {\endinnercustomgeneric}
}

\newcustomtheorem{customthm}{Theorem}
\newcustomtheorem{customlem}{Lemma}

\makeatletter
\def\old@comma{,}
\catcode`\,=13
\def,{%
  \ifmmode%
    \old@comma\discretionary{}{}{}%
  \else%
    \old@comma%
  \fi%
}
\makeatother

\newcommand\numberthis{\addtocounter{equation}{1}\tag{\theequation}}

\usepackage{xcolor}
\definecolor{darkblue}{rgb}{0.1,0.1,0.8}
\definecolor{DarkGreen}{rgb}{0,0.6,0}
\definecolor{brickred}{rgb}{0.8, 0.25, 0.33}
\definecolor{britishracinggreen}{rgb}{0.0, 0.26, 0.15}
\definecolor{calpolypomonagreen}{rgb}{0.12, 0.3, 0.17}
\definecolor{ao(english)}{rgb}{0.0, 0.5, 0.0}
	\definecolor{cadmiumgreen}{rgb}{0.0, 0.42, 0.24}
\definecolor{burgundy}{rgb}{0.5, 0.0, 0.13}
\usepackage{etoolbox}

\providetoggle{Blue_revision}
\settoggle{Blue_revision}{true}

\providetoggle{Track}
\settoggle{Track}{true}
\newcommand{\addv}[3]{%
	\iftoggle{Track}{%
    	\IfEqCase{#1}{%
       	 	{a}{\ifthenelse{\equal{#2}{ON}}{{\color{cadmiumgreen}#3}}{#3}}%
        	{b}{\ifthenelse{\equal{#2}{ON}}{{\color{brickred}#3}}{#3}}%
       		{c}{\ifthenelse{\equal{#2}{ON}}{{\color{burgundy}#3}}{#3}}%
    	}[\PackageError{tree}{Undefined option to tree: #1}{}]%
	}{#3}%
}

 \usepackage{hyperref}

\hypersetup{
colorlinks=true, %
  pdfstartview={FitH},
    linkcolor=red,
    citecolor=blue,
    urlcolor={blue!80!black}
}

\usepackage{graphicx}

\newcounter{relctr} 
\everydisplay\expandafter{\the\everydisplay\setcounter{relctr}{0}} 

\AtBeginDocument{} 

\global\long\def\NN{\mathbb{N}}

\global\long\def\EE{\mathbb{E}}
\global\long\def\PP{\mathbb{P}}

\global\long\def\11{\mathbbm{1}}

\newcommand{\CP}{\mathcal{P}}

\newcommand{\CX}{\mathcal{X}}
\newcommand{\CY}{\mathcal{Y}}

\global\long\def\+{\oplus}

\newcommand{\prob}[1]{\PP\Big\{  #1 \Big\} }
\def\<{\langle}
\def\>{\rangle}

\ifdefined \var 
  \renewcommand{\var}{\mathsf{var}}
\else
  \newcommand{\var}{\mathsf{var}}
\fi

\ifdefined \abs \else
 \newcommand{\abs}[1]{\lvert#1\rvert}
\fi

\ifdefined \norm \else
 \newcommand{\norm}[1]{\lVert#1\rVert}
\fi

\ifdefined \set 
  \renewcommand{\set}[1]{\left\{#1\right\}}
\else
  \newcommand{\set}[1]{\left\{#1\right\}}
\fi

\newcommand*{\medcup}{\mathbin{\scalebox{1}{\ensuremath{\bigcup}}}}%

\DeclareMathOperator*{\argmax}{arg\,max}

\usepackage{stackengine}
\def\deq{\mathrel{\ensurestackMath{\stackon[1pt]{=}{\scriptstyle\Delta}}}}

\def\etal{\textit{et al.}}

\providecommand{\tr}{tr}

\ifdefined \Tr 
  \renewcommand{\Tr}[1]{\tr \Big\{#1\Big\}}
\else
  \newcommand{\Tr}[1]{\tr \Big\{#1\Big\}}
\fi

\def\PMAC{\mathcal{P}}
\def\CAVC{\mathcal{C}_F^{VLC}}

\def\Tleps{{\tau_{\epsilon}}}
\def\Tueps{{\tau^{\epsilon}}}

\def\Tuepstld{\Tueps}
\def\Tuepstldi{\Tueps}
\def\Tlepstldi{\Tleps}
\def\Zprune{L_n}

\def\J{{J}}
\def\D{{D}}
\def\Qbar{{\bar{Q}_r}}

\def\xstar{x^*_{r}}
\def\Xstar{\xstar} 

\def\wstar{w^*_{r}}
\def\Wstar{\wstar}

\def\sr{}
\usepackage[nolist,nohyperlinks]{acronym}

\newacro{ptp}[PtP]{Point-to-Point}

\newacro{iid}[i.i.d.]{independent and identically distributed} 
\newacro{IID}[i.i.d.]{independent and identically distributed} 
\newacro{DMC}[DMC]{discrete memoryless channel}
\acrodefplural{DMC}{discrete memoryless channels}

\newacro{VLC}[VLC]{variable length code}
\acrodefplural{VLC}{variable length codes}

\newacro{wrt}[w.r.t]{with respect to}
\newacro{AVC}[AVC]{arbitrary varying channel}
\acrodefplural{AVC}{arbitrary varying channels}

\settoggle{Track}{false}

\newcommand{\addva}[1]{\addv{a}{off}{#1}}

\newcommand{\addvb}[1]{\addv{b}{off}{#1}}

\begin{document}
\title{Upper Bounds on the Feedback Error Exponent of Channels With States and  Memory}

\author{ Mohsen Heidari \IEEEauthorrefmark{1}, Achilleas Anastasopoulos \IEEEauthorrefmark{2} and  S.\ Sandeep Pradhan \IEEEauthorrefmark{2},\\
       \IEEEauthorrefmark{1}  CS Dept., Purdue University, West Lafayette, IN. \\
      \IEEEauthorrefmark{2} 
EECS Dept., University of  Michigan, Ann Arbor, MI. \\
\IEEEauthorrefmark{1}\tt mheidari@purdue.edu,
 \IEEEauthorrefmark{2}\tt \{anastas, pradhanv\}@umich.edu
   \thanks{This work was supported by NSF grant CCF-2132815.}   }
\maketitle
\begin{abstract}
As a class of state-dependent channels, Markov channels have been long studied in information theory for characterizing the feedback capacity and error exponent. This paper studies a more general variant of such channels where the state evolves via a general stochastic process, not necessarily Markov or ergodic. The states are assumed to be unknown to the transmitter and the receiver, but the underlying probability distributions are known. For this setup, we derive an upper bound on the feedback error exponent and the feedback capacity with variable-length codes. The bounds are expressed in terms of the directed mutual information and directed relative entropy. The bounds on the error exponent are simplified to Burnashev's expression for discrete memoryless channels. Our method relies on tools from  the theory of martingales to analyze a stochastic process defined based on the entropy of
the message given the past channel's outputs.  
\end{abstract}

\section{Introduction}
 Communications over channels with feedback has been a longstanding problem in information theory literature. The early works on \acp{DMC} pointed to negative answer as to whether feedback can increase the capacity \cite{Shannon_zero}. Feedback,
though, improves the channel's error exponent --- the maximum attainable exponential rate of decay of the error probability. 
The improvements are obtained using \acp{VLC}, where the communication length depends on the channel's relizations.   In a seminal work, Burnashev \cite{Burnashev} completely characterized the error exponent of \acp{DMC}  with noiseless and causal feedback. This characterization has a simple,  yet intuitive, form:
\begin{align}\label{eq: E(R) error exponent }
E(R)=C_1(1-\frac{R}{C}) ,
\end{align}
where $R$ is the (average) rate  of transmission, $C$ is the capacity of the channel, and $C_1$ is the maximum exponent for binary hypothesis testing over the channel. It is equal to  the maximal relative entropy between conditional output distributions. The Burnashev's exponent can significantly exceed the sphere-packing exponent, for no-feedback communications, as it approaches capacity with nonzero slope. The use of VLCs is shown to be essential to establish these resutls, as no improvements is gained using fixed-length codes \cite{Dobrushin,Haroutunian,Sheverdyaev1982}. 

This result led to the question as to whether the
feedback improves capacity or error exponent of more general channels, modeling non-traditional communications involving memory and intersymbol interference (ISI). Among such models are channels with states where the transition probability of the channel varies depending on its state which itself evolves based on the past inputs and state realizations. Depending on the variants of this formulation, the agents may have no knowledge about the state (e.g. arbitrarily varying channels) or the may exactly know the state \cite{Goldsmith1996}.  When state is known at the transmitter and the receiver,  feedback can improve the error exponent.  Particularly, Como, et al, \cite{Como2009} extended Burnashev-type exponent to finite-state ergodic Markov channels with known state and derived a similar form as in \eqref{eq: E(R) error exponent }, under some ergodicity assumptions. The error exponent for channels with more general state evolution is still unknown. Only the feedback capacity of such channels when restricted to fixed-length codes is known \cite{Tatikonda2009}.

 This papers studies the feedback error exponent for channels with more general state evolution and allowing VLCs. More precisely, we study discrete channels with  states where the state evolves as an arbitrary stochastic process (not necessarily ergodic or Markov) depending on the past realizations.  Furthermore, the realization of the states are assumed to be unknown but the transmitter or the receiver may know the underlying probability distribution governing the evolution of the state.  However, noiseless output is available at the transmitter with one unite of delay. The main contributions are two fold. First, we prove an upper bound on the error exponent of such channels which has the familiar form 
\begin{align*}
E(R)\leq  \sup_{N>0}\sup_{P^N\in \CP^N}D(P^N)(1-\frac{R}{I(P^N)}), 
\end{align*}
where $D$ is the directed relative entropy, $I$ is the directed mutual information, and $\CP^N$ is a collection of ``feasible" probability distributions. As a special case, the bound simplifies to the Burnashev's expression when the channel is DMC. Second, we introduce an upper bound on the feedback capacity of \acp{VLC} for communications over these channels with stochastic states.  This upper bound generalizes the results of Tatikonda and Mitter \cite{Tatikonda2009}, and Purmuter \etal \cite{Permuter2009}  where fixed-length codes are studied.  Our approach relies on analysis of the entropy of the stochastic process defined based on entropy of the message given the past channel's output. We analyze the drift of the entropy via tools from the theory of martingales.

Related works on the capacity and error exponent of channels with feedback are extensive. Starting with DMCs with fedback, Yamamoto and Itoh \cite{Yamamoto} introduced a  two-phase iterative for achieving the Burnashev exponent. Also, error exponent of DMCs with feedback and cost constraints is studied in \cite{Nakiboglu2008}. Also channels with state and feedback has been studied under various frameworks on the evolution model of the sates and whether they are known at the transmitter or the receiver. On one exterem of such models are arbitrarily varying channels  \cite{Ahlswede1997}. The feedback capacity these channels for fixed-length codes is derived in \cite{Tatikonda2009}.   Tchamkerten and Telatar \cite{Tchamkerten2005} studied the universality of Burnashev error exponent. They considered communication setups where the parties have no  exact knowledge of the statistics of the channel but know it belongs to a certain class of DMCs. The authors proved that no zero-rate coding scheme
achieves the Burnashev's exponent simultaneously for all the DMC's in the class. However, they showed positive results  for two families of such channels (e.g., binary symmetric  and Z) \cite{Tchamkerten2006}. Another class of channels with state are Markov channels that has been studied extensively for deriving their capacity \cite{Goldsmith1996,Chen2005,Bae2010} and error exponent using fixed-length codes \cite{Tatikonda2009}. A lower bound on the error exponent of unifilar channels is derived \cite{Anastasopoulos2017}, where the states is a deterministic function of the previous ones.  
Other variants of this problem have been studied, including continuous-alphabet channels \cite{Horstein1963,Schalkwijk-Kailath}, and  multi-user channels \cite{Kramer-thesis,Heidari2018}.

\section{Problem Formulation and Definitions}\label{sec: problem formulation}
The formal definitions  are presented in this section. For short hand, we use $[1:M]$ to denote $\set{1,2, ..., M}.$

A discrete channel with stochastic state has three finite sets $\CX, \CY$, and  $\mathcal{S}$  representing the input, output, and state of the channel, respectively. Consider a collection of channels $\mathcal{Q}:=\set{Q(\cdot|\cdot, s): s\in \mathcal{S}}$, indexed by $s\in\mathcal{S}$, where each element $Q(\cdot|\cdot, s):\CX\rightarrow \CP(\CY)$ is the transition probability of the channel at state $s$. 
The states $\{S_t\}_{t>0}$, evolve according to a conditional probability distribution $P_{t, S} (s_{t} | s^{t-1}, x^{t-1}), t>0$ depending on the past inputs and state realizations.  As a result, after $t$ uses of the channel with $x^{t-1}, s^{t-1}, y^{t-1}$ being the channels input, state and output, the next output is given by 
\begin{align*}
P(s_{t}, y_{t} | x^{t-1}, s^{t-1}, y^{t-1}) = P_{t, S} (s_{t} | s^{t-1}, x^{t-1}) Q(y_{t}| x_{t}, s_{t}).
\end{align*}
Such evolution of the states induces memory over the time as it depends on past inputs. 

After each use of the channel, the output of the channel $y_t$ is available at the transmitter with one unit of delay. Moreover, we allow \acp{VLC} for communications, where where both the transmitter and the receiver do not know the state of the channel. More precisely, the setup is defined as follows.
\begin{definition}
An $(M, N)$-VLC for communications over a channel $\mathcal{Q}$ with states and feedback is defined  by 
\begin{itemize}
\item A message $W$ with uniform distribution over  $[1 : M]$.


\item Encoding functions $e_{t}: [1: M] \times \mathcal{Y}^{t-1}\rightarrow \mathcal{X},   t\in \NN.$ 
 
\item Decoding functions $d_t: \mathcal{Y}^t \rightarrow [1 : M],  t\in \NN.$
\item  A stopping time $T$ \ac{wrt} the filtration $\mathcal{F}_t$ defined as the $\sigma$-algebra of $Y^t$ for $t\in \NN$. \addva{Furthermore, it is assumed that $T$ is almost surely bounded as $T \leq N$.}
\end{itemize}

\end{definition}
For technical reasons, we study a class of $(M, N)$-VLCs for which the parameter $N$ grows sub-exponentially with $\log M,$ that is $N\leq (\log M)^m$ for some fixed number $m$. An example is the sequence $(M^{(n)}, N^{(n)})$-VLCs, $n\geq 1,$ where $M^{(n)} = 2^{nr_1},  N^{(n)} \leq n^m,$
with $r_1,r_2, m>0$ being fixed parameters .

In what follows, for any $(M, N)$-VLC, we define average rate, error probability, and error exponent. Given a message $W$, the $t$-th output of the transmitter is denoted by $X_{t}=e_{t}(W, Y^{t-1})$, where $Y^{t-1}$ is the noiseless feedback upto time $t$. Let $\hat{W}_t=d_t(Y^t)$ represent the estimate of the decoder about the message. Then, at the end of the stopping time $T$, the decoder declares $\hat{W}_T$ as the decoded message. The average rate and (average) probability of error for a VLC are defined as 
\begin{align*}
R \deq \frac{\log_2 M}{\EE[T]}, \quad P_e \deq \prob{\hat{W}_T \neq W}.
\end{align*}
 
\begin{definition}\label{def:capcity}
A rate $R$ is achievable for a given channel with stochastic states, if there exists a sequence of $(M^{(n)}, N^{(n)})$-VLCs such that
 \begin{align*}
 \limsup_{n\rightarrow \infty} P^{(n)}_e = 0, \qquad \limsup_{n\rightarrow \infty}\frac{\log M^{(n)}}{\EE[T^{(n)}]} \geq R, 
 \end{align*}
 and $N^{(n)}\leq (n)^m, \forall n>1$, where $m$ is fixed. The feedback capacity, $\CAVC$, is the convex closure of all achievable rates.
\end{definition}
Naturally, the error exponent of a VLC with probability of error $P_e$ and stopping time $T$ is defined as $E \deq -\frac{\log_2 P_e}{\EE[T]}$. The following definition formalizes this notion.
\begin{definition}\label{def:ErrExponent}
An error exponent function $E(R)$ is said to be achievable for a given channel, if for any rate $R>0$  there exists a sequence of $(M^{(n)}, N^{(n)})$-VLCs  such that
 \begin{align*}
\liminf_{n\rightarrow \infty} -\frac{\log P^{(n)}_e}{\EE[T^{(n)}]} &\geq E(R),~~  & \limsup_{n\rightarrow \infty}\frac{\log M^{(n)}}{\EE[T^{(n)}]} &\geq R, 
 \end{align*}
 and $ \limsup_{n\rightarrow \infty} M^{(n)} = \infty$ with $N^{(n)}\leq (n)^m, \forall n>1$, where $m$ is fixed. The reliability function is the supremum of all achievable reliability functions $E(R)$. 
\end{definition}

\section{Main Results}
We start with deriving an upper bound on the feedback capacity of channels with stochastic states and allowing VLCs. The expressions are based on the \textit{directed information}  as  introduced in  \cite{Massey1990} and  defined as 
\begin{align}\label{eq:directed MI}
I({X}^n \rightarrow {Y}^n) \deq \sum_{i=1}^n I(X_{i}; Y_i| Y^{i-1}).
\end{align}
We further extend this notion to variable-length sequences. Consider a stochastic process $\{(X_t, Y_t)\}_{t>0}$ and let  $T$ be a  (bounded) stopping time w.r.t an induced filtration $\mathcal{F}_t, t>0$. Then, the directed mutual information is defined as   
\begin{align}\label{eq:directed MI T}
I({X}^T \rightarrow {Y}^T) \deq \EE\big[ \sum_{t=1}^T I({X}_t; {Y}_t~ \|\mathcal{F}_{t-1})\big].  
\end{align}

Now, we are ready for an upper bound on the feedback capacity. For any integer $N$, let $\CP^N$ be the set of all $N$-letter distributions  $P_{X^N, S^N, Y^N}$ on $\mathcal{X}^N\times \mathcal{S}^N\times \mathcal{Y}^N$ that factor as  
\begin{align}\label{eq: factorization of P}
 \prod_{\ell=1}^N P_{\ell, X}(x_{\ell}| x^{\ell-1}, y^{\ell-1})P_{\ell, S}(s_{\ell}| s^{\ell-1}, x^{\ell-1})Q(y_\ell|x_{\ell}, s_{\ell}).
\end{align}

 Next, we have the following result on the capacity with the proof in Appendix \ref{proof:thm: VLC Capacity}. 

\begin{theorem}\label{thm:VLC Capacity}
The feedback capacity of a channel with stochastic states is bounded as
\begin{align*}
\CAVC\leq \sup_{N>0}\sup_{P^N\in \CP^N}\sup_{T: T\leq N} \frac{1}{\EE[T]} I(X^T \rightarrow Y^T ),
\end{align*}
where $T$ is a stopping time with respect to $\mathcal{F}_t, t>0$.
\end{theorem}
Observe that for a trivial stopping time $T=N$, the bound simplifies to that for fixed-length codes as given in\cite{Tatikonda2009}. 

%

\subsection{Upper Bound on the Error Exponent}
We need a notation to proceed. Consider a pair of random sequences $(X^n,Y^n)\sim P_{X^nY^n}$. Let $X_r^*$ be the MAP estimation of $X_r$ from observation $Y^{r-1}$, that is $X_r^* =\argmax_{x} \PP\set{X_r = x | Y^{r-1} = y^{r-1}}$.
Also,  let $\Qbar = P_{Y_r|X_{r}, Y^{r-1}}$ which is the effective channel (averaged over possible states) from the transmitter's perspective at time $r$.
With this notation, we define the directed KL-divergence as 
\begin{align*}
D(X^{n} \rightarrow Y^n)  &\deq \max_{x^n}\sum_{r=1}^n  D_{KL}\Big(\Qbar(\cdot |X_r^*, Y^{r-1}) \big\| \Qbar(\cdot |x_r, Y^{r-1}) ~\big|~ Y^{r-1}\Big).  
\end{align*}
Intuitively, $D(X^{n} \rightarrow Y^n) $ measures the sum of the expected ``distance" between the channels probability distribution conditioned on the MAP symbol versus the worst symbol, across different times $r\in [1:n]$. 

\begin{theorem}\label{thm:err exp}
The error exponent of a channel with stochastic states is bounded as 
\begin{align*}
E(R) \leq \sup_{N\in \NN} \sup_{P^N \in \CP^N} \sup_{\substack{ T: T\leq N}} \sup_{\substack{ T_1: T_1\leq T} } D(P^N) \Big(1 - \frac{R}{I(P^N)}\Big),
\end{align*}
where $T,T_1$ are stopping times, and 
\begin{align*}
I(P^N) &= \frac{1}{\EE[T_1]}    I(X^{T_1} \rightarrow Y^{T_1}),\\
D(P^N) &=\frac{1}{\EE[T-T_1]} D(X_{T_1+1}^{T} \rightarrow Y_{T_1+1}^{T}). 
\end{align*}
\end{theorem}
In the next section, we present our proof techniques.

\section{Proof of Theorem \ref{thm:err exp}}
\begin{figure}[ht]
\centering
\includegraphics[scale=1.1]{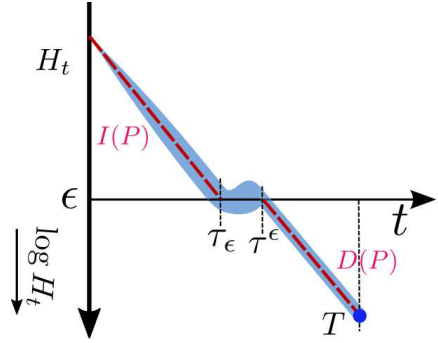} 
\caption{Entropy drifts over $t$. The first part till $\Tleps$, as in \eqref{eq:Tleps}, is the linear drift with the expected slope $I(P)$ (dashed line). From $\Tleps$ to $\Tueps$, as in \eqref{eq:Tueps}, are the fluctuations around $\epsilon$ (shaded region). Then, from $\Tueps$ to $T$ is the logarithmic drift ($\log H_t$) with the expected slope of $D(P)$ (the second dashed line).}
\label{fig:drift}
\end{figure}

The proof follows by a careful study of the drift of the entropy of the message $W$ conditioned on the channel's output at each time $t$. Define the following random process: 
\begin{align}\label{eq:H processes}
H_t & = H(W |\mathcal{F}_t ), t>0,
\end{align}
where $\mathcal{F}_t$ is  the $\sigma$-algebra of $Y^t$. We show that $H_t$ drifts in three phases: (i)  linear drift (data phase) until reaching a small value ($\epsilon$); (ii) fluctuation phase with values around $\epsilon$; and (iii)  logarithmic drift (hypothesis testing phase) till the end. We derive bounds on the expected slope of the drifts and prove that the length of the fluctuation phase is asymptotically negligible as compared to the overall communication length ( Fig. \ref{fig:drift}). 

More precisely, we have the following argument by defining a pruned time random process $\set{t_n}_{n>0}$.
First, for any $\epsilon \geq 0$ and $N\in \NN$  define the following random variables
\begin{align}\label{eq:Tleps}
\Tleps &\deq \inf\set{t>0: H_t\leq \epsilon} \wedge N\\\label{eq:Tueps}
\Tueps &\deq \sup\set{t>0: H_{t-1}\geq \epsilon} \wedge N
\end{align}
Then the pruned time process is defined as 
\begin{align}\label{eq:tn}
t_n \deq \begin{cases}
               n & \text{if}~ n< \Tleps\\
               n\vee \Tueps & \text{if}~ \Tleps \leq n\leq N\\
                N  & \text{if}~ n> N                 
            \end{cases}
\end{align} 
Note that $\Tleps$ is a stopping time with respect to $\{H_t\}_{t>0}$ but this is not the case for $\Tueps$.

\begin{lem}\label{lem:pruned sub martingale 1}
Suppose a non-negative random process $\set{H_r}_{r>0}$ has the following properties w.r.t a filtration $\mathcal{F}_r, r>0,$
\begin{subequations}
\begin{alignat}{4}\label{eq:H linear K1}
 \EE[H_{r+1} - H_r |\mathcal{F}_r]&\geq -k_{1,r+1},  & \qquad \text{if}~  H_r & \geq \epsilon, \\\label{eq:H log K2}
\EE[\log H_{r+1} - \log H_r |\mathcal{F}_r]&\geq -k_{2,r+1}      & \text{if}~   H_r & < \epsilon\\\label{eq:H log K3}
\abs{\log H_{r+1} - \log H_r} & \leq k_3    & &\\\label{eq:H K4}
\abs{ H_{r+1}-H_r} &\leq k_4   & &
\end{alignat}
\end{subequations}
where $k_{1,r}, k_{2,r}, k_3, k_4$ are non-negative numbers and $k_{1,r}\leq k_{2,r}$ for all $r>0$.  Given $\epsilon \in (0,1)$, and $D\geq I>0$, let 
\begin{align*}
Z_t &\deq \frac{H_t-\epsilon}{I}\11\{H_t\geq \epsilon\}+\big(\frac{\log \frac{H_t}{\epsilon}}{D}+f(\log \frac{H_t}{\epsilon}) \big) \11\{H_t<\epsilon\},
\end{align*}
where $f(y)=\frac{1-e^{\lambda y}}{\lambda D}$ with $\lambda >0$. Further define $\set{S_t}_{t>0}$ as 
\begin{align*}
S_t &\deq \sum_{r=1}^{t\wedge \Tleps} \frac{k_{1,r}}{I} + \sum_{r=t\wedge \Tleps+1}^{t\wedge \Tueps} \frac{k_4}{I} \11\{H_{r-1} \geq \sqrt{\epsilon}\}+\sum_{r=t\wedge \Tueps+1}^{t} \frac{k_{2,r}}{D} + \sqrt{\epsilon} \frac{N}{I}\11\{t\geq \Tueps\}.
\end{align*}
Let $\set{t_n}_{n>0}$ be as in \eqref{eq:tn} but w.r.t  $\set{H_r}_{r>0}$. Lastly define the random process $\set{\Zprune}_{n>0}$ as $\Zprune \deq Z_{t_n}+S_{t_n}.$ Then, for small enough $\lambda>0$ the process $\set{\Zprune}_{n>0}$ is a sub-martingale with respect to the time pruned filtration $\mathcal{F}_{t_n}, n>0$. 
\end{lem}

\begin{IEEEproof}
The objective is to prove $\EE[L_{n+1}- L_n|y^{t_n}] \geq 0$ almost surely for all $n\geq 1$ and $y^{t_n}$. We prove the lemma by considering three cases depending on $n$. 

\noindent\textbf{Case (a). $n< \Tleps -1$:} From the definition of $t_n$ in \eqref{eq:tn}, in this case $t_n=n$ and $t_{n+1} =n+1.$ Also, as the time did not reach $\Tleps$, then $H_n>\epsilon$ and $H_{n+1}>\epsilon$. Therefore, in this case, the random process of interest equals to 
\begin{align}\nonumber
L_n &= Z_{t_n} + S_{t_n}= Z_n+S_n = \frac{H_n-\epsilon}{I} + \sum_{r=1}^{n} \frac{k_{1,r}}{I} \\\nonumber
L_{n+1} &= Z_{t_{n+1}} + S_{t_{n+1}}= Z_{n+1}+S_{n+1}\\\label{eq:case a 1}
& = \frac{H_{n+1}-\epsilon}{I} + \sum_{r=1}^{n+1} \frac{k_{1,r}}{I} . 
\end{align}
As a result, the difference between $L_n$ and $L_{n+1}$ satisfies the following
\begin{align*}
\EE[(L_{n+1}- L_n)&\11\{n< \Tleps -1\}|y^{t_n}]\\
 &= \EE[(L_{n+1}- L_n)\11\{n< \Tleps -1\}|y^{n}]\\
&= \EE[L_{n+1}- L_n|y^{n}]\11\{n< \Tleps -1\},
\end{align*}
where the first equality holds as $t_n=n$ and the second equality holds as $\Tleps$ is a stopping time which implies that $\11\{n< \Tleps -1\}$ is a function of $y^n$. Next, from \eqref{eq:case a 1}, the difference term above is bounded as 
\begin{align*}
\EE[L_{n+1}- L_n|y^{n}] & =  \EE[\frac{H_{n+1}-H_n}{I} + \frac{k_{1,n+1}}{I} | y^n ]\\
& = \frac{\EE[ H_{n+1}-H_n|y^n ] }{I} + \frac{k_{1,n+1}}{I} \geq 0,
\end{align*}
where the last inequality follows from \eqref{eq:H linear K1}. As a result, we proved that $\EE[(L_{n+1}- L_n)\11\{n< \Tleps -1\}|y^{t_n}] \geq 0$. 

\noindent\textbf{Case (b). $n = \Tleps -1$:} In this case, $t_n = n$ implying that $H_n>\epsilon$ and $t_{n+1} = (n+1) \vee \Tueps$. Furthermore, since, $n+1 = \Tleps \leq \Tueps$, then $t_{n+1} = \Tueps$. Consequently, the random process equals to 
\begin{align*}
L_n &= Z_{n} + S_{n} = \frac{H_n-\epsilon}{I} + \sum_{r=1}^{n} \frac{k_{1,r}}{I} \\
L_{n+1} &= Z_{\Tueps} + S_{\Tueps}= (\frac{H_\Tueps-\epsilon}{I})\11\{H_\Tueps\geq \epsilon\}\\
&~~+\big(\frac{\log H_\Tueps - \log \epsilon}{D}+f(\log \frac{H_\Tueps}{\epsilon}) \big) \11\{H_\Tueps<\epsilon\}\\
&~~ + \sum_{r=1}^{\Tleps} \frac{k_{1,r}}{I} + \sum_{r=\Tleps+1}^{\Tueps} \frac{k_4}{I} \11\{H_{r-1} \geq \addvb{\sqrt{\epsilon}}\} +\addvb{\sqrt{\epsilon}} \frac{N}{I}. 
 \end{align*} 
 Note that $Z_{\Tueps}$ does not necessarily equal to the logarithmic part. The reason is that $\Tueps$ is pruned by $N$ as in \eqref{eq:Tueps}. Thus, $H_{\Tueps}$ can be greater than $\epsilon$ when $\Tueps = N$. We proceed by bounding $Z_{\Tueps}$. Note that, for small enough $\lambda$ the following inequality holds 
\begin{equation}\label{eq:fy 1}
\frac{\epsilon}{I}(e^{y}-1)-\frac{y}{D} < f(y), \qquad   -k_3 < y <0 .
\end{equation}
 Applying inequality \eqref{eq:fy 1} with $y = \log \frac{H_\Tueps}{\epsilon}$, we can write that
 \begin{align}\nonumber
 Z_{\Tueps} &> (\frac{H_\Tueps-\epsilon}{I})\11\{H_\Tueps\geq \epsilon\}+\big(\frac{H_\Tueps-\epsilon}{I} \big)\11\{H_\Tueps<\epsilon\}\\\label{eq:}
 & = \frac{H_\Tueps-\epsilon}{I}
   \end{align}  
Consequently, the difference $L_{n+1}-L_n$ satisfies the following 
\begin{align*}
\EE[&(L_{n+1}- L_n)\11\{n= \Tleps -1\}|y^{t_n}]\\
 &= \EE[L_{n+1}- L_n|y^{n}]\11\{n= \Tleps -1\}\\\numberthis \label{eq:case c 1}
&\geq \EE\bigg[ \frac{H_{\Tueps}-H_n }{I} +  \frac{k_{1,\Tleps}}{I} + \sum_{r=\Tleps+1}^{\Tueps} \frac{k_4}{I} \11\{H_{r-1} \geq \sqrt{\epsilon}\} +\sqrt{\epsilon}\frac{N}{I}\Big| y^n \bigg]\11\{n= \Tleps -1\}
\end{align*}

Next, we bound the first term above as  
\begin{align*}
H_{\Tueps}-H_n  &= H_{n+1}-H_n  + \sum_{r=n+2}^{\Tueps} (H_{r}-H_{r-1}), 
\end{align*}
where in the first equality, we add and subtract the intermediate terms $H_r, n+1\leq r \leq \Tueps-1$. Next,we substitute the above terms in the right-hand side of \eqref{eq:case c 1}. As $n+2 = \Tleps+1$, then we obtain that 
\begin{align*}
\eqref{eq:case c 1} &=  \EE\bigg[ \frac{H_{n+1}-H_n }{I} +  \frac{k_{1,\Tleps}}{I} + \sum_{r=\Tleps+1}^{\Tueps} \Big( \frac{H_{r}-H_{r-1}}{I}+ \frac{k_4}{I} \11\{H_{r-1} \geq \addvb{\sqrt{\epsilon}}\}\Big)+\sqrt{\epsilon}\frac{N}{I}\Big| y^n \bigg]\11\{n= \Tleps -1\}\\\numberthis \label{eq:case c 2}
&\geq   \EE\bigg[\sum_{r=\Tleps+1}^{\Tueps} \Big( \frac{{H_{r}-H_{r-1}}}{I}+\frac{k_4}{I} \11\{H_{r-1} \geq \sqrt{\epsilon}\}\Big)+ \sqrt{\epsilon}\frac{N}{I}\Big| y^n \bigg]\11\{n= \Tleps -1\},
\end{align*}
where the inequality holds from \eqref{eq:H linear K1} and the fact that $n+1=\Tleps$. Next, by factoring $I$ and the indicator function inside the expectation, we have the following chain of inequalities 
\begin{align*}
\eqref{eq:case c 2} & = \frac{1}{I}\EE\bigg[\sum_{r=\Tleps+1}^{\Tueps} \Big((H_{r}-H_{r-1})+{k_4}\Big) \11\{H_{r-1} \geq \sqrt{\epsilon}\}+  \Big( ({H_{r}-H_{r-1}}) \11\{H_{r-1} < \sqrt{\epsilon}\}\Big) +\sqrt{\epsilon} N\Big| y^n \bigg]\11\{n= \Tleps -1\}\\
&\stackrel{(a)}{\geq}  \frac{1}{I}\EE\bigg[\sum_{r=\Tleps+1}^{\Tueps} \Big( ({H_{r}-H_{r-1}}) \11\{H_{r-1} <\sqrt{\epsilon}\}\Big) + \sqrt{\epsilon} N\Big| y^n \bigg]\11\{n= \Tleps -1\}\\
&\stackrel{(b)}{\geq}  \frac{1}{I}\EE\bigg[\Big(\sum_{r=\Tleps+1}^{\Tueps} -H_{r-1} \11\{H_{r-1} < \sqrt{\epsilon}\}\Big) +  \sqrt{\epsilon} N\Big| y^n \bigg]\11\{n= \Tleps -1\}\\
&\stackrel{(c)}{>}  \frac{1}{I}\EE\left[\Big(\sum_{r=\Tleps+1}^{\Tueps} -\addvb{\sqrt{\epsilon}} \Big) +\addvb{\sqrt{\epsilon}} N \Big| y^n \right]\11\{n= \Tleps -1\}\\
&\stackrel{(d)}{\geq}  \frac{1}{I}\EE\left[\Big(\sum_{r=1}^{N} -\addvb{\sqrt{\epsilon}} \Big) +\addvb{\sqrt{\epsilon}} N\Big| y^n \right]\11\{n= \Tleps -1\}\\
&\geq 0,
\end{align*}
where (a) is due to \eqref{eq:H K4}, inequality (b) holds as $H_r\geq 0$, inequality (c) holds as  $H_{r-1} \11\{H_{r-1} < \epsilon\}<\epsilon$, and lastly (d) holds as $\Tueps \leq N$. To sum up, we proved that 
\begin{align*}
\EE[(L_{n+1}- L_n)\11\{n= \Tleps -1\}|y^{t_n}] \geq 0.
\end{align*}

\noindent\textbf{Case (c). $n \geq \Tleps$:} This is the last case. Note that if
$n<\Tueps$, then $t_n =t_{n+1} = \Tueps$. Thus, immediately, $L_{n+1} -L_n =0$ almost surely.   Otherwise, if $n\geq \Tueps$ and $\Tueps =N$ or if  $n\geq N$, then $t_n=t_{n+1}=N$ and hence   $L_{n+1} -L_n =0$. Therefore, it remains to consider the case that $\Tueps<N$ and $\Tueps\leq n<N$. Therefore, $t_n = n$ and $t_{n+1} = n+1$. Furthermore, as $n+1>n\geq \Tueps$ and $\Tueps<N$, then $H_n<\epsilon$ and $H_{n+1}<\epsilon$, implying that we are in the logarithmic drift. Therefore, we have that
\begin{align*}
 L_n = Z_n & = \frac{\log H_n -\epsilon}{D} +f(\log \frac{H_n}{\epsilon})+S_n\\
 L_{n+1} = Z_{n+1} & = \frac{\log H_{n+1} -\epsilon}{D} +f(\log \frac{H_{n+1}}{\epsilon})+S_{n+1}. 
 \end{align*} 
Hence, to sum up the above sub-cases, we conclude that when $n \geq \Tleps$, then 
\begin{align*}
L_{n+1}-L_n &= \frac{\log H_{t_{n+1}} - \log H_{t_{n}}}{D}+f(\log \frac{H_{t_{n+1}}}{\epsilon})-f(\log \frac{H_{t_{n}}}{\epsilon})+S_{t_{n+1}}-S_{t_n}.
\end{align*}
Note that from \eqref{eq:H log K2}, the following inequality holds
\begin{align*}
\EE\left[\frac{\log H_{t_{n+1}} - \log H_{t_{n}}}{D}+S_{t_{n+1}}-S_{t_n} \Big| y^{t_n}\right] \geq 0.
\end{align*}
Therefore, the difference $L_{n+1}-L_n$ satisfies the following 
\begin{align*}
\EE[(L_{n+1}&- L_n)\11\{n \geq \Tleps\}|y^{t_n}] &\\
& =\EE[(L_{n+1}- L_n)|y^{t_n}]\11\{n \geq \Tleps\}\\
& \geq  \EE\left[f(\log \frac{H_{t_{n+1}}}{\epsilon})-f(\log \frac{H_{t_{n}}}{\epsilon})\Big| y^{t_n}\right] \11\{n \geq \Tleps\}.
\end{align*}
Next, we provide an argument similar to \ac{ptp} case. That is, we use the Taylor's theorem for $f$. We only need to consider the case that $\Tueps<N$ and $\Tueps\leq n<N$ implying that $t_n = n$ and $t_{n+1} = n+1$.   Using the Taylor's theorem we can write
\begin{align*}
&f(\log \frac{H_{{n+1}}}{\epsilon})= f(\log \frac{H_{{n}}}{\epsilon})+\frac{\partial f}{\partial y}\Big|_{y = \log \frac{H_{{n}}}{\epsilon}} \big(\log H_{n+1} - \log H_n \big)+\frac{\partial^2 f}{\partial y^2}\Big|_{y = \zeta}(\log \frac{H_{n+1}}{H_n})^2,
\end{align*}
where $\zeta$ is between $\log \frac{H_{n+1}}{\epsilon}$ and $\log \frac{H_n}{\epsilon}$ and 
\begin{align*}
\frac{\partial f}{\partial y}\Big|_{y = \log \frac{H_{{n}}}{\epsilon}}  = -\frac{e^{\lambda \log \frac{H_n}{\epsilon}}}{I}, \qquad \frac{\partial^2 f}{\partial y^2}\Big|_{y = \zeta} = -\frac{\lambda }{I}e^{\lambda \zeta}.
\end{align*}
As a result, we have that
\begin{align*}
&\EE\left[f(\log \frac{H_{{n+1}}}{\epsilon})-f(\log \frac{H_{{n}}}{\epsilon})\Big| y^{n}\right]\\
 & = \EE\left[-\frac{e^{\lambda \log \frac{H_n}{\epsilon}}}{I} \big(\log \frac{H_{n+1}}{H_n} \big)-\frac{\lambda }{I}e^{\lambda \zeta}(\log \frac{H_{n+1}}{H_n})^2\Big| y^n \right]\\
&= \EE\left[-\frac{e^{\lambda \log \frac{H_n}{\epsilon}}}{I} \big(\log \frac{H_{n+1}}{H_n} \big)-\frac{\lambda }{I}e^{\lambda ( \zeta \pm \log \frac{H_n}{\epsilon})}(\log \frac{H_{n+1}}{H_n})^2\Big| y^n \right]\\
&\stackrel{(a)}{\geq}  \EE\left[-\frac{e^{\lambda \log \frac{H_n}{\epsilon}}}{I} \big(\log \frac{H_{n+1}}{H_n} \big)-\frac{\lambda }{I}e^{\lambda ( k_3 + \log \frac{H_n}{\epsilon})}(\log \frac{H_{n+1}}{H_n})^2\Big| y^n \right]\\
&\geq  \EE\left[-\frac{e^{\lambda \log \frac{H_n}{\epsilon}}}{I} \big(\log \frac{H_{n+1}}{H_n} \big)-\frac{\lambda k_3^2}{I}e^{\lambda ( k_3 + \log \frac{H_n}{\epsilon})}\Big| y^n \right]\\
&=  -\frac{e^{\lambda \log \frac{H_n}{\epsilon}}}{I} \EE[\log \frac{H_{n+1}}{H_n} | y^n ]-\frac{\lambda k_3^2}{I}e^{\lambda ( k_3 + \log \frac{H_n}{\epsilon})}\\
&=  -\frac{e^{\lambda \log \frac{H_n}{\epsilon}}}{I} \underbrace{\EE[\log \frac{H_{n+1}}{H_n} | y^n ]}_{\leq k_3}-\frac{\lambda k_3^2e^{\lambda k_3} }{I}e^{\lambda \log \frac{H_n}{\epsilon}}\\
&\geq   \frac{e^{\lambda \log \frac{H_n}{\epsilon}}}{I} k_3  -\frac{\lambda k_3^2e^{\lambda k_3} }{I}e^{\lambda \log \frac{H_n}{\epsilon}}\\
&=   \big( \frac{k_3}{I}  -\frac{\lambda k_3^2e^{\lambda k_3} }{I}\big) e^{\lambda \log \frac{H_n}{\epsilon}}\\
& \geq 0,
\end{align*}
where inequality (a) holds as $\abs{\zeta - \log \frac{H_n}{\epsilon}} \leq \abs{\log \frac{H_{n+1}}{\epsilon} - \log \frac{H_n}{\epsilon}}\leq k_3$. The last inequality holds for sufficiently small $\lambda >0$.

Lastly, combining all cases from (a) to (c), we prove that $\EE[L_{n+1}-L_n | y^n]\geq 0$ which completes the proof.
\end{IEEEproof}

Now, we show that $\set{H_t}_{t>0}$ as in \eqref{eq:H processes} has the conditions in Lemma \ref{lem:pruned sub martingale 1}. First \eqref{eq:H linear K1} holds because of the following lemma.
\begin{lem}\label{lem: linear drift}
Given any $(M, N)$-VLC,  the following inequality holds almost surely for $1 \leq r \leq N$
\begin{align}\label{eq:linear drift J}
\EE[H_{r}-H_{r-1}|\mathcal{F}_{r-1}] & = - \J_{r},
\end{align}
where $\J_r \deq I(X_{r}; Y_r | \mathcal{F}_{r-1})$ with the induced $P_{X^N,Y^N}\in \CP^N$.
\end{lem}
\begin{IEEEproof}
For any $y^{r-1}$, we have that
\begin{align*}
\EE[H_{r}-&H_{r-1}| y^{r-1}]  = H(W|Y_r, y^{r-1}) -H(W| y^{r-1})\\
& = -I(W; Y_r | y^{r-1})\\ 
& = -I(W, X_{r}; Y_r |y^{r-1})\\
& = -H(Y_r | y^{r-1})+ H(Y_r | W, X_{r}, y^{r-1})\\
&= -H(Y_r |  y^{r-1})+ H(Y_r | X_{r})\\
& = - J_r.
\end{align*} 
Hence the lemma is proved.
\end{IEEEproof}
Condition \eqref{eq:H log K2} holds as a result of the following lemma that is given in Appendix \ref{app:proof:lem:log}.
\begin{lem}\label{lem: log drift}
For any $(M, N)$-VLC and $\epsilon \in [0,\frac{1}{2}]$, if $H_r<\epsilon$, then the following inequality holds almost surely
\begin{align}\label{eq:log drift D}
\EE[\log H_{r}-\log H_{r-1}|\sr \mathcal{F}_{r-1}] & \geq - \D_{r}+O(h_b^{-1}(\epsilon)), 
\end{align}
where and  $\D_r$ is a function of $y^{r-1}$ and is defined as 
\begin{align}\label{eq:D_r definition}
\D_r \deq \max_{x\in \mathcal{X}}D_{KL}\Big(\Qbar(\cdot | \xstar, y^{r-1})~ \|~ \Qbar(\cdot| x, y^{r-1})\Big),
\end{align}
where $\Qbar = P_{Y_r|X_{r}, Y^{r-1}}$ is the average channel from the transmitter's perspective, and  $x^*_{r}$ is the MAP input symbol given by $x^*_{r}= \argmax_{x} \PP\set{X = x | Y^{r-1} =y^{r-1}}$. 
\end{lem}
Condition \eqref{eq:H log K3} is a direct consequence of Lemma 4 in \cite{Burnashev}:
\begin{remark}\label{rem: H_t+1 - H_tis bounded}
If $Q(\cdot |\cdot ,\cdot)$ are positive everywhere then   $\abs{\log H_r-\log H_{r-1}} \leq \eta$, where 
\begin{align*}
\eta \deq \max_{x_1, x_2 \in\mathcal{X}}\max_{s_1, s_2\in\mathcal{S}}\max_{y\in \mathcal{Y}} \log \frac{Q(y|x_1, s_1)}{Q(y|x_2, s_2)}.
\end{align*}
\end{remark}
Lastly, \eqref{eq:H K4} holds as $H_r\leq \log M$ which implies that 
\begin{align*}
\abs{H_r-H_{r-1}} \leq \max \{ H_r, H_{r-1} \} \leq \log M.
\end{align*}
Thus, we apply Lemma \ref{lem:pruned sub martingale 1} on $\{H_t\}_{t>0}$ with
\begin{align*}
k_{1,r} = \J_r, ~ k_{2,r} = \D_r, ~ k_3 = \eta, ~ k_4 = \log M,
\end{align*}
and constants $I,D$ to be specified later. Therefore, $\{L_n\}_{n >0}$ as in the lemma is a sub-martingale w.r.t $\mathcal{F}_{t_n}, n>0$.

\subsection{Connection to the error exponent}
Since $\{L_n\}_{n >0}$ is a sub-martingale, then $L_0\leq \EE[L_{T\vee \Tuepstldi}]$, where $T$ is the stopping time used in the VLC and $\Tueps$ is as in \eqref{eq:Tueps}.  Note that $L_0 = \frac{\log M}{I}$. In what follows, we analyze $\EE[L_{T\vee \Tuepstldi}]$. 

By definition $L_n = Z_{t_n}+S_{t_n}$. Since, $T\leq N$, then from \eqref{eq:tn} we have that $t_{T\vee \Tuepstldi} = (T\vee \Tuepstldi) \vee \Tuepstldi=T\vee\Tuepstldi$. Therefore,  
\begin{align*}
\frac{\log M}{I} & \leq  \EE[L_{T\vee \Tuepstldi}]\\
& =  \EE\Big[\frac{H_{T\vee \Tuepstldi}-\epsilon}{I}\11\{H_{T\vee \Tuepstldi}\geq \epsilon\} +\big(\frac{\log (H_{T\vee \Tuepstldi}/\epsilon)}{D}+ f(\log \frac{H_{T\vee \Tuepstldi}}{\epsilon}) \big) \11\{H_{T\vee \Tuepstldi}<\epsilon\}+ S_{T\vee \Tuepstldi}\Big]\\
&\stackrel{(a)}{\leq} \EE\Big[\frac{H_{T\vee \Tuepstldi}+\epsilon}{I}+\frac{\log (H_{T\vee \Tuepstldi}/ \epsilon)}{D}+f(\log \frac{H_{T\vee \Tuepstldi}}{\epsilon})\Big]+\EE\big[S_{T\vee \Tuepstldi}\big]\\
&\stackrel{(b)}{\leq} \EE\left[\frac{H_{T\vee \Tuepstldi}+\epsilon}{I}\right]+\EE\left[\frac{\log H_{T\vee \Tuepstldi} - \log \epsilon}{D}\right]+\frac{1}{\lambda D}+ \EE\big[S_{T\vee \Tuepstldi}\big]\\\numberthis \label{eq:error exp up 1 }
& \stackrel{(c)}{\leq}\frac{\EE\big[H_{T\vee \Tuepstldi}\big]+\epsilon}{I} +\frac{\log \EE[H_{T\vee \Tuepstldi}]- \log \epsilon}{D}+\frac{1}{\lambda D} + \EE\big[S_{T\vee \Tuepstldi}\big]
\end{align*}
where $(a)$ follows by changing $-\epsilon$ to $+\epsilon$ for the linear part and from the following inequality for the logarithmic part
\begin{align*}
(\log x - \log \epsilon) \11\{x < \epsilon\} &\leq \log x - \log \epsilon.
\end{align*}
Inequality (b) and (c) follow from Jensen's inequality, concavity of $\log(x)$ and the inequality $f(y)\leq \frac{1}{\lambda D}$. 

Next, we bound $\EE\big[H_{T\vee \Tuepstldi}\big]$ in \eqref{eq:err exponent up eq 1}. As conditioning reduces the entropy, then 
\begin{align*}
H_{T\vee \Tuepstldi} = H(W | Y^{T\vee \Tuepstldi}) \leq  H(W | Y^{T})=H_T,
\end{align*}
where the inequality holds as $Y^T$ is a function of $Y^{T\vee \Tuepstldi}$.  Next, Fano's inequality implies that  
\begin{align}\label{eq:Fanos}
\EE\big[H_{T\vee \Tuepstldi}\big]\leq  \EE\big[H_{T}\big]  = \EE[H(W | Y^T)]  \leq \alpha(P_e),
\end{align}
where $\alpha(P_e)=h_b(P_e) + P_e \log(M)$ is the Fano's expression.  Therefore, from 
\eqref{eq:Fanos}, we obtain that
\begin{align*}
\frac{\log M}{I} &\leq \frac{\alpha(P_e)+\epsilon}{I} +\frac{\log \alpha(P_e) - \log \epsilon}{D}+\frac{1}{\lambda D} + \EE\big[S_{T\vee \Tuepstldi}\big].
\end{align*}
Rearranging the terms gives the following inequality
\begin{align*}
\frac{-\log \alpha(P_e)}{D} &\leq \frac{\alpha(P_e)+\epsilon}{I} +\frac{- \log \epsilon}{D}+\frac{1}{\lambda D}+ \EE\big[S_{T\vee \Tuepstldi}\big] - \frac{\log M}{I}.
\end{align*}
Therefore, multiplying by $D$ and dividing by $\EE[T]$ give the following
\begin{align}\label{eq:err exp alpha}
\frac{-\log \alpha(P_e)}{\EE[T]} \leq D \Big( \frac{ \EE\big[S_{T\vee \Tuepstldi}\big]}{\EE[T]} - \frac{R}{I}\Big)+U(P_e, M, \epsilon),
\end{align}
where we used the fact that $\frac{\log M}{\EE[T]} \geq R$, and that
\begin{align}\label{eq:Ui}
U(P_e, M, \epsilon) =  R\big( \frac{\alpha(P_e)+\epsilon}{I \log M} +\frac{- \log \epsilon}{D\log M}+\frac{1}{\lambda D\log M}\big).
\end{align}
Next, for the left hand side of \eqref{eq:err exp alpha}, we can write that 
\begin{align*}
&-\log \alpha(P_e) = -\log P_e -\log \frac{\alpha(P_e) }{P_e}\\
&= -\log P_e -\log \Big(-\log P_e -\frac{(1-P_e)}{P_e} \log (1-P_e) + \log M   \Big)\\
& \geq -\log P_e -\log \Big(-\log P_e -\frac{1}{P_e} \log (1-P_e) + \log M \Big)\\
& \geq -\log P_e -\log \Big(-\log P_e +2 + \log M \Big),
\end{align*}
where the last inequality follows because  $\log(x)\geq 1-\frac{1}{x}$ for $x>0$ implying that  $\log (1-P_e) \geq 1-\frac{1}{1-P_e} = \frac{-P_e}{1-P_e}$; and hence, $-\frac{1}{P_e} \log (1-P_e) \leq \frac{1}{1-P_e}\leq 2$ as $P_e \leq \frac{1}{2}$. Therefore, by factoring $-\log P_e$ we have that 
\begin{align*}
-\log \alpha(P_e) \geq (-\log P_e) ( 1 -\Delta ),
\end{align*}
where 
\begin{align}\label{eq:Delta}
\Delta = \frac{\log \big(-\log P_e +2+ \log M \big)}{-\log P_e}.
\end{align}
Therefore, from \eqref{eq:err exp alpha}  we get the following bound on the error exponent
\begin{align}\label{eq:err exponent up eq 1}
\frac{-\log P_e}{\EE[T]} &\leq \frac{D}{1-\Delta}\Big( \frac{ \EE\big[S_{T\vee \Tuepstldi}\big]}{\EE[T]} - \frac{R}{I}+U(P_e, M, \epsilon)\Big).
\end{align}

Next, we find appropriate $I$ and $D$ so that $\EE\big[S_{T\vee \Tuepstldi}\big] \approx \EE[T]$. Further, we show that $\Delta$ and $U(P_e, M, \epsilon)$ converge to zero for any sequence of VLCs satisfying Definition \ref{def:ErrExponent}.

We proceed with the following lemma that is proved in Appendix \ref{app:proof:lem:bounding S}.

\begin{lem}\label{lem:bounding S}
Given $\epsilon > \alpha(P_e)$ and with 
\begin{align*}
I = \frac{1}{\EE[\Tlepstldi]} \EE\Big[ \sum_{r=1}^{\Tlepstldi}\J_r \Big], \qquad D = \frac{1}{\EE[T -\Tlepstldi]} \EE\Big[ \sum_{r=\Tlepstldi+1}^{T}\D_r \Big],
\end{align*}
the inequality $\EE\big[S_{T\vee \Tuepstldi}\big]\leq \EE[T] (1+ V(\epsilon, N))$ holds, where $V(\epsilon,N) =\frac{R_i}{I} \big( \addvb{\sqrt{\epsilon}} N \big)+ \addvb{\sqrt{\epsilon}} \frac{N}{\EE[T] I}$.
\end{lem}

Therefore, with \eqref{eq:err exponent up eq 1}, we get the desired upper bound by appropriately setting $I$ and $D$ as in the lemma. Hence, we get
\begin{equation}\label{eq:err exponent up eq 2}
\frac{-\log P_e}{\EE[T]} \leq  \frac{D}{1-\Delta}\Big(1 - \frac{R}{I}+U(P_e, M, \epsilon)+  V(\epsilon,N)\Big).
\end{equation}
We show that for any $(M^{(n)}, N^{(n)})$-VLCs as in Definition \ref{def:ErrExponent} the residual terms $U,V, \Delta$ converge to zero as $n\rightarrow \infty$. 
It is easy to see that $\Delta$ as in \eqref{eq:Delta} converges to zero as $P_e^{(n)}\rightarrow 0$. Further, by setting $\epsilon^{(n)} = (\frac{1}{N^{(n)}})^3$, we can check that $\lim_{n\rightarrow \infty} V(\epsilon^{(n)}, N^{(n)})=0$. It remains to show the convergence of $U(\cdot)$ as in \eqref{eq:Ui}. The convergence of the third term in \eqref{eq:Ui} follows as $\lim_{n\rightarrow \infty}\frac{1}{M^{(n)}}= 0$. For the second term, as  $\epsilon^{(n)} = (\frac{1}{N^{(n)}})^3$ then we have that
\begin{align*}
\lim_{n\rightarrow \infty} \frac{-\log \epsilon^{(n)}}{\log M^{(n)}} = \lim_{n\rightarrow \infty} \frac{3 \log N^{(n)} }{\log M^{(n)}} =0,
 \end{align*} 
where the last equality holds as $N^{(n)}$ grows sub-exponentially with $n$. The convergence of the first term also follows from the fact that $\lim_{n\rightarrow \infty}\alpha(P_e^{(n)})=0$, as $P_e^{(n)}$ converges exponentially fast\footnote{The exponential convergence of $P_e^{(n)}$ holds because otherwise the error exponent is zero.}. 

Hence, by maximizing over all distributions and from Definition \ref{def:ErrExponent}, we get the desired upper bound on the error exponent.
\begin{align}\label{eq: Err Exp up 1}
\limsup_{n\rightarrow \infty} -\frac{\log P^{(n)}_e}{\EE[T^{(n)}]} \leq \sup_{N\in \NN} \sup_{P^N \in \PMAC^N} \sup_{T:  T\leq N} \left\{D(P^N) \Big(1 - \frac{R}{I(P^N)}\Big)\right\}, 
\end{align}
where the maximizations are taken over all distributions $P^N\in \PMAC^N$. 
Further,  $I(P^N)$ and $D(P^N)$ are defined as in Lemma \ref{lem:bounding S} with the distribution $P^N$. 

\section*{Conclusion}
This paper presents an upper bound on the feedback error exponent and feedback capacity  of channels with stochastic states, where the states evolve according to a general stochastic process. The results are based on the analysis of the drift of the entropy of the message as a random process.

\bibliographystyle{IEEEtran}
\bibliography{main_arxiv}
\newpage
\onecolumn
\appendices
\section{Proof of Theorem \ref{thm:VLC Capacity}}\label{proof:thm: VLC Capacity}
From Definition \ref{def:capcity}, consider an achievable rate  $R$. Based on the definition of achievability, consider any $(M, N)$-VLC with the probability of error $P_e$ and a stopping time $T$ that is less than $N$ almost surely such that $R \leq \frac{\log M}{\EE[T]}$. 

From the definition of $H_{t}$ in \eqref{eq:H processes}, we can write
\begin{align*}
\EE[H_{T}] &= \sum_{t \geq 0} \PP\{T=t\}  \EE\big[H(W| \mathcal{F}_t)| T=t\big]\\
& =  \sum_{t \geq 0} \PP\{T=t\}  H\big(W|Y^t, T=t\big)\\
& = H(W|Y^T),
\end{align*}
where $Y^T$ is a random variable taking values from a subset in $\mathcal{Y}^N$.  From Fano's inequality as in \eqref{eq:Fanos}, we can bound the above quantities as $\EE\big[H_{T}\big] \leq \alpha(P_e),$ where $\alpha(P_e)= h_b(P_e) + P_e \log M.$ Next, we start with bounding the rate $R$. Since at time $t=0$, the message $W$ has the uniform distribution, then we have that   
\begin{align*}
\log M &= H(W) = I(W; Y^T) + H(W|Y^T)\\\numberthis\label{eq: bound on M1}
&\leq I(W; Y^T) + \alpha(P_e).
\end{align*}
We proceed by showing that 
\begin{align*}
I(W; Y^T) \leq I(X^T \rightarrow Y^T).
\end{align*}

We first pad $Y^T$ to make it a sequence of length $N$. Let $\xi$ be an auxiliary symbol and define 
$$Y^N =(Y_1, Y_2, \cdots, Y_T, \xi, \xi, \cdots, \xi).$$
Similarly, we extend the encoding functions and the channel's transition probability to include $\xi$. Specifically, after the stopping time $T$, the encoders send the constant symbol $\xi$ and the channel outputs $\xi$ to the receiver. More precisely,  
\begin{align*}
e(W, Y^n) = \xi, ~~~\forall n\geq T, \qquad Q(y|\xi, s) =  \11\set{y=\xi}, ~~~ \forall x, x, y.
\end{align*}
This auxiliary adjustment is only for  tractability of the analysis as it does not affect the performance of the code. Specifically, the mutual information stays the same by replacing $Y^T$ with $Y^N$:
\begin{align*}
 I(W; Y^N) &= I(W; Y^T) + I(W; Y^N | Y^T) \\
& =  I(W; Y^T ) + I(W; \xi_{T+1}^N |  Y^T) \\\numberthis \label{eq:Mu eq 1}
& = I(W; Y^T).
 \end{align*} 
From the chain rule, we have that
\begin{align*}
I(W; Y^N)& = \sum_{r=1}^N I(W; Y_r| Y^{r-1})\\
& = \sum_{r=1}^N H(Y_r | Y^{r-1}) - H(Y_r |W, Y^{r-1})\\
& = \sum_{r=1}^N H(Y_r | Y^{r-1}) - H(Y_r |W, X_{r}, Y^{r-1})\\
& = \sum_{r=1}^N H(Y_r |Y^{r-1}) - H(Y_r |X_{r}, Y^{r-1})\\
& = \sum_{r=1}^N I(X_{1,r}; Y_r |  Y^{r-1})\\\numberthis\label{eq:Mu eq 2}
&= I(X^N \rightarrow Y^N ).
\end{align*}

Next, we show that the directed mutual information above equals to the following: 
\begin{align*}
I(X^N \rightarrow Y^N) =  I(X^T \rightarrow Y^T \|~ X^T) = \EE\Big[\sum_{r=1}^T I(X_{r}; Y_r |\mathcal{F}_{r-1})\Big],
\end{align*}
where the second equality is due to the definition given in \eqref{eq:directed MI T}.
Note that $I(X_{1,r}; Y_r | \mathcal{F}_{r-1}) = 0$ almost surely for any $r>T$ as $Y_r=\xi$. Therefore, we have that
\begin{align*}
 I(X^T \rightarrow Y^T ) & =  \EE\Big[\sum_{r=1}^N I(X_{r}; Y_r |  \mathcal{F}_{r-1})\Big]\\
 & =\sum_{r=1}^N  \EE\big[I(X_{r}; Y_r |  \mathcal{F}_{r-1})\big]\\
 & =\sum_{r=1}^N  I(X_{r}; Y_r |  Y^{r-1})\\\numberthis\label{eq:Mu eq 3}
 & = I(X^N \rightarrow Y^N \|).
\end{align*}
Therefore, combining \eqref{eq: bound on M1}-\eqref{eq:Mu eq 3} gives an upper bound on $\log M$. Dividing both sides by $\EE[T]$ gives the following upper bound on $R$
\begin{align}\nonumber
R &\leq \frac{\log M}{\EE[T]}\leq  \frac{1}{\EE[T]}I(X^T \rightarrow Y^T)+\frac{\alpha(P_e)}{\EE[T]}\\\label{eq:R1 up}
&\leq \sup_{N>0}\sup_{P^N\in \CP^N}\sup_{{\text{stop time}} T: T\leq N} \frac{1}{\EE[T]} I(X^T \rightarrow Y^T )+\frac{\alpha(P_e)}{\EE[T]}
\end{align}
The first term above is the desired expression. The second term is vanishing as $P_e \rightarrow 0$. Hence the proof is complete. 

\section{Proof of Lemma \ref{lem: log drift}}\label{app:proof:lem:log}
\begin{IEEEproof}
 Define the following quantities
 \begin{subequations}
\begin{alignat}{3}\label{eq:mu_w}
\mu(w)&=\prob{W=w| Y^{r-1}=y^{r-1}}\\\label{eq:mu_w_yr}
\mu(w, y_{r})&=\prob{W=w| Y^{r-1}=y^{r-1},Y_{r}=y_{r}}\\\label{eq:Q_w}
Q_{w}(y_{r})&=\prob{Y_{r}=y_{r}| W=w, Y^{r-1}=y^{r-1}},
\end{alignat}
 \end{subequations}
where $i\in [1:M], y_{t+1}\in \mathcal{Y}$. Let $\Wstar \in [1: M]$ be the most likely message condition on $y^{r-1}$. That is $\wstar = \max_{w\in [1: M]}\mu(w)$.
First, we show that having $H_{r-1} < \epsilon, \epsilon\in [0,1)$, we conclude that $\mu(\wstar)= 1-\eta_1(\epsilon)$ with $\eta_1$ being a function satisfying $\lim_{\epsilon \rightarrow 0} \eta_1(\epsilon) =0.$ The argument is as follows:
 
Using the grouping axiom we have
\begin{align}\label{eq:Htld eq 1}
H_{r-1}=H(W| y^{r-1})=h_b( \mu(\wstar))+(1-\mu(\wstar))H(\hat{W}),
\end{align}
where $\hat{W}$ is a random variable with probability distribution $P(\hat{W}=w)=\frac{\mu(w)}{1-\mu(\wstar)}, ~w\in [1:M], w\neq \wstar$. Hence, having $H_{r-1}\leq \epsilon$ implies that $h_b(\mu(\wstar))\leq \epsilon$.
Taking the inverse image of $h_b$ implies that either $\mu(\wstar)\geq 1-h_b^{-1}(\epsilon)$ or $\mu(\wstar)\leq h_b^{-1}(\epsilon)$, where $h_b^{-1}:[0,1]\rightarrow [0,\frac{1}{2}]$ is the lower-half inverse function of $h_b$.  We show that the second  case is not feasible. For this purpose, we show that the inequality $\mu(\wstar)\leq h_b^{-1}(\epsilon)\leq \frac{1}{2}$ implies that $H_{r-1}\geq 1$ which is a contradiction  with the original assumption $H_{r-1} \leq \epsilon < 1$. This statement is proved in the following proposition. With this argument, we conclude that $H_{r-1}\leq \epsilon$ implies that $\mu(\wstar)\geq 1-\eta_1(\epsilon)$, where $\eta_1(\cdot) = h_b^{-1}(\cdot)$.

\begin{proposition}
Let $W$ be a random variable taking values from a finite set $\mathcal{W}$. Suppose that $P_W(w)\leq \frac{1}{2}$ for all $w\in \mathcal{W}$. Then $H(W)\geq 1$. 
\end{proposition}
\begin{IEEEproof}
The proof follows from an induction on $|\mathcal{W}|$. For $|\mathcal{W}|=2$ the condition in the statement implies that $W$ has uniform distribution and hence $H(W)=1$ trivially. Suppose the statement holds for $|\mathcal{W}|=n-1$. Then, we prove it for $|\mathcal{W}|=n$.  Sort elements of $\mathcal{W}$ in an descending order according to $P_W(w)$, from the most likely (denoted by $W$) to the least likely ($w_n$). If $P_W(w_n)=0$, then the statement holds trivially from the induction's hypothesis.  Suppose $P_W(w_n)>0$. In this case, we can reduce  $H(W)$ by increasing $P_W(W)$ and decreasing $P_W(w_n)$ so that $P_W(W)+P_W(w_n)$ remains constant. In that case, either $P_W(w_n)$ becomes zero or $P_W(W)$ reaches the limit $\frac{1}{2}$. The first case happens if $P_W(W)+P_W(w_n)\leq \frac{1}{2}$. For that, the statement $H(W)\geq 1$ follows from the induction's hypothesis, as there are only $(n-1)$ elements with non-zero probability.  It remains to consider the second case in which $P_W(W)=\frac{1}{2}$ and $P_W(w_n)>0$. Again, we can further reduce the entropy by increasing $P_W(w_2)$ and decreasing $P_W(w_n)$ while $P_W(w_2)+P_W(w_n)$ remains constant. Observe that $P_W(w_2)+P_W(w_n)\leq \frac{1}{2}$ as $\sum_{i=2}^n P_W(w_i) = 1-P_W(W)=\frac{1}{2}$ and $p_W(w_i)>0$.  Hence, after this redistribution process $P_W(w_n)$ becomes zero. Then, the statement $H(W)\geq 1$ follows from the induction's hypothesis, as there are only $(n-1)$ elements with non-zero probability.
\end{IEEEproof}


We proceed with the proof of the lemma by applying Lemma 7 in \cite{Burnashev}:
\begin{customlem}{7}[\cite{Burnashev}]
For any non-negative sequence of numbers $p_\ell,\mu_i$ and $\beta_{i,l}, \ell \in [1:L], i\in [1:N]$ the following inequality holds
\begin{align*}
\sum_{\ell=1}^L p_\ell \log\Big( \frac{\sum_{i=1}^N \mu_i}{\sum_{i=1}^N \beta_{i,\ell}}  \Big) \leq \max_{i} \sum_{\ell=1}^L p_{\ell} \log \frac{\mu_i}{\beta_{i,\ell}}.
\end{align*}
\end{customlem}

As a result,
\begin{align*}
\EE[&\log H_{r-1}-\log H_{r}|\sr \mathcal{F}_{r-1}] \\
&=\sum_{y_r} P(y_r|\sr y^{r-1}) \log\Big(\frac{-\sum_{w} \mu(w)\log \mu(w) }{-\sum_{w}\mu(w,y_r)\log \mu(w,y_r)}\Big)\\
& \leq  \max_{w} \Gamma(w),
\end{align*}
where 
\begin{align*}
\Gamma(w) = \sum_{y_r} P(y_r|\sr y^{r-1}) \log \frac{-\mu(w)\log \mu(w)}{-\mu(w,y_r)\log \mu(w,y_r)}.
\end{align*}

Note that 
\begin{align}\label{eq:1-mu1 approximate}
\mu(w, y_{r})&=\frac{\mu(w) Q_w(y_r)}{P(y_r|y^{r-1})},
\end{align}
Therefore, for a fixed $w\neq \wstar$ we have that
\begin{align*}
&-\Gamma(w)\\
 &= \sum_{y_r} P(y_r|\sr y^{r-1}) \log\bigg[ \frac{Q_w(y_r)}{ P(y_r|y^{r-1})}\Big(1+\frac{\log\frac{P(y_r|y^{r-1})}{Q_w(y_r)}}{-\log \mu(w)}\Big)\bigg]\\
&=\sum_{y_r} P(y_r|\sr y^{r-1}) \log \frac{Q_w(y_r)}{ P(y_r|y^{r-1})}+ \log\Big(1+\frac{\log\frac{P(y_r|y^{r-1})}{Q_w(y_r)}}{-\log \mu(w)}\Big)\\
&\stackrel{(a)}{=} -D_{KL}\Big( P(\cdot|y^{r-1})~\|~ Q_w \Big)\\
&~~ + \sum_{y_r} P(y_r|y^{r-1}) \log\Big(1+\frac{\log\frac{P(y_r|y^{r-1})}{Q_w(y_r)}}{-\log \mu(w)}\Big)
\end{align*}
The summation in the last equality is bounded using the inequality $\log (1+x)\geq x$ for all $x\geq -1$. Hence we get that 
\begin{align*}
-\Gamma(w)&\geq  -D_{KL}\Big( P(\cdot|y^{r-1})~\|~ Q_w \Big)\\
&~~ + \sum_{y_r} P(y_r|y^{r-1})\frac{\log\frac{P(y_r|y^{r-1})}{Q_w(y_r)}}{-\log \mu(w)}\\
&= -D_{KL}\Big( P(\cdot|y^{r-1})~\|~ Q_w \Big)\Big(1-\frac{1}{-\log \mu(w)}\Big) 
\end{align*}

Having $\mu(w)\leq \eta_1(\epsilon)$ for all $w\neq \wstar$, we have that 
\begin{align}\nonumber
\Gamma(w) &\leq D_{KL}\Big( P(\cdot|y^{r-1})~\|~ Q_w \Big)\Big(1-\frac{1}{-\log \eta_1(\epsilon)}\Big)\\\label{eq:gamma w}
&\leq D_{KL}\Big( P(\cdot|y^{r-1})~\|~ Q_w \Big),
\end{align}
where the last inequality follows as $-\log \eta_1(\epsilon)\leq 0$.

Next we consider the case $w=\wstar$. We use the Taylor's theorem for the function $f(x) = -x\log x$ around $x=1$. With that, $f(x) = (1-x) - \zeta (1-x)^2$ for some $\zeta$ between $x$ and $1$. Hence, with $x= \mu(\Wstar)$, we have that 
\begin{align*}
-\mu(\Wstar) \log \mu(\Wstar) &\leq (1-\mu(\wstar))-\zeta (1-\mu(\wstar))^2\\
& \leq (1-\mu(\wstar)). 
\end{align*}
Next, from the inequality $\log x \leq x-1, \forall x>0$, we have that 
\begin{align*}
-\mu(\Wstar,y_r) \log \mu(\Wstar,y_r) \geq \mu(\Wstar,y_r) (1- \mu(\Wstar,y_r)).
\end{align*}
As a result of these inequalities, we have that 
\begin{align*}
\Gamma(\wstar) &= \sum_{y_r} P(y_r|y^{r-1}) \log \frac{-\mu(\wstar)\log \mu(\wstar)}{-\mu(\wstar,y_r)\log \mu(\wstar,y_r)}\\
&\leq \sum_{y_r} P(y_r|y^{r-1}) \log \frac{(1-\mu(\wstar))}{\mu(\Wstar,y_r) (1- \mu(\Wstar,y_r))}\\
&=  \sum_{y_r} P(y_r|y^{r-1}) \log \frac{1-\mu(\wstar)}{1- \mu(\Wstar,y_r)}\\\numberthis \label{eq:wstar eq1}
&\quad  - \sum_{y_r} P(y_r|y^{r-1}) \log \mu(\Wstar,y_r). 
\end{align*}

We proceed with simplifying the first summation above. From \eqref{eq:1-mu1 approximate}, we have that 
\begin{equation}\label{eq:mu wstar and y}
\mu(\Wstar, y_{r}) =\frac{\mu(\Wstar)Q_{\Wstar}(y_{r})}{P(y_r|y^{r-1}) }.
\end{equation}
 Therefore,
\begin{align}\label{eq:1-mu1 approximate}
(1-\mu(\Wstar, y_{r}))&=(1-\mu(\Wstar)) \frac{\sum_{w\notin \Wstar} \frac{\mu(w)}{1-\mu(\Wstar)}   Q_{w}(y_{r})}{P(y_r|y^{r-1})}.
\end{align}
Thus, the first summation on the right-hand side of \eqref{eq:wstar eq1} is simplified as 
\begin{align*}
&\sum_{y_r} P(y_r|y^{r-1}) \log \frac{P(y_r|y^{r-1})}{\sum_{w\notin \Wstar} \frac{\mu(w)}{1-\mu(\Wstar)}   Q_{w}(y_{r})}\\
& = D_{KL}\Big( P(\cdot|y^{r-1})~\|~ Q_{\sim \wstar} \Big),
\end{align*}
where $Q_{\sim \wstar}(y_r)= \sum_{w\notin \Wstar} \frac{\mu(w)}{1-\mu(\Wstar)}   Q_{w}(y_{r})$ for all $y_r\in \mathcal{Y}$.
Next, we bound the second summation in \eqref{eq:wstar eq1}. Using \eqref{eq:mu wstar and y}, we have that 
\begin{align*}
- \sum_{y_r} & P(y_r|y^{r-1}) \log \mu(\Wstar,y_r)\\
  &= -\log \mu(\wstar) - \sum_{y_r} P(y_r|y^{r-1}) \log \frac{Q_{\Wstar}(y_{r})}{P(y_r|y^{r-1}) }\\
 &= -\log \mu(\wstar) + D_{KL}\Big( P(\cdot|y^{r-1})~\|~ Q_{\wstar} \Big)\\\numberthis\label{eq:wstar eq2}
 &\leq 2\eta_1(\epsilon)+ D_{KL}\Big( P(\cdot|y^{r-1})~\|~ Q_{\wstar} \Big),
\end{align*}
where the last inequality holds from the fact that $-\log (1-x) \leq \frac{x}{1-x}$ and that $\mu(\wstar)\geq 1-\eta_1(\epsilon)$, implying $-\log \mu(\wstar)\leq -\log (1-\eta_1(\epsilon)) \leq \frac{\eta_1(\epsilon)}{1-\eta_1(\epsilon)}\leq 2\eta_1(\epsilon)$ which holds as $\eta_1(\epsilon)\leq \frac{1}{2}$. Note that $P(\cdot|y^{r-1}) = \mu(\wstar)Q_{\wstar} + (1-\mu(\wstar))Q_{\sim \wstar}$. Therefore, from the convexity of the relative entropy the right-hand side of \eqref{eq:wstar eq2} is bounded by 
\begin{align*}
\eqref{eq:wstar eq2}&\leq 2\eta_1(\epsilon) + \mu(\wstar) D_{KL}\Big( Q_{\wstar} ~\|~ Q_{\wstar} \Big)\\
&\quad +(1-\mu(\wstar)) D_{KL}\Big( Q_{\sim \wstar} ~\|~ Q_{\wstar} \Big)\\
&\leq 2\eta_1(\epsilon)+ \eta_1(\epsilon) D_{KL}\Big( Q_{\sim \wstar} ~\|~ Q_{\wstar} \Big)\\
& \leq (2+d^{\max}) \eta_1(\epsilon),
\end{align*}
where the last inequality is from the definition of $d^{\max}$ being the maximum relative entropy of the channel. As a result of the above argument, we have that 
\begin{align*}
\Gamma(\wstar) &\leq D_{KL}\Big( P(\cdot|y^{r-1})~\|~ Q_{\sim \wstar} \Big) + (2+d^{\max}) \eta_1(\epsilon)\\
&\stackrel{(a)}{\leq} \sum_{w\notin \Wstar} \frac{\mu(w)}{1-\mu(\Wstar)}  D_{KL}\Big( P(\cdot|y^{r-1})~\|~ Q_{w} \Big)\\
&\quad + (2+d^{\max}) \eta_1(\epsilon)\\\numberthis\label{eq:gamma wstar}
&\stackrel{(b)}{\leq} \max_{w\neq \wstar}D_{KL}\Big( P(\cdot|y^{r-1})~\|~ Q_{w} \Big) + (2+d^{\max}) \eta_1(\epsilon)
\end{align*}
where (a) is due to the convexity of the relative entropy and the definition of $Q_{\sim \wstar}$. Inequality (b) follows as $\frac{\mu(w)}{1-\mu(\Wstar)}$ form a probability distribution on $w\neq \wstar$.

Note that the right-hand side of \eqref{eq:gamma w} and \eqref{eq:gamma wstar} depends on the messages. In what follows we remove this dependency. Note that the convexity of the relative entropy gives 
\begin{align*}
D_{KL}\Big( P(\cdot|y^{r-1})~\|~ Q_{w} \Big) &\leq \mu(\wstar) D_{KL}\Big( Q_{\wstar}~\|~ Q_{w} \Big)\\
&\quad  + (1-\mu(\wstar)) D_{KL}\Big( Q_{\sim \wstar}~\|~ Q_{w} \Big)\\
&\leq D_{KL}\Big( Q_{\wstar}~\|~ Q_{w} \Big)+\eta_1(\epsilon) d^{\max}.
\end{align*}

As a result the bound in  \eqref{eq:gamma wstar} is simplified to the following
\begin{align}
\Gamma(\wstar) &\leq \max_{w} D_{KL}\Big( Q_{\wstar}~\|~ Q_{w} \Big) + 2(1+d^{\max}) \eta_1(\epsilon)
\end{align}
Similarly, the bound on $\Gamma(w)$ in \eqref{eq:gamma w} is simplified to 
\begin{align}
\Gamma(w) &\leq D_{KL}\Big( Q_{\wstar}~\|~ Q_{w} \Big)+\eta_1(\epsilon) d^{\max}
\end{align}

Combining the two bounds above, we finally can bound the logarithmic drift as 
\begin{align*}
\EE[&\log H_{r-1}-\log H_{r}|\mathcal{F}_{r-1}] \leq \max_{w} \Gamma(w)\\
&\leq \max_{w} D_{KL}\Big( Q_{\wstar}~\|~ Q_{w} \Big) + 2(1+d^{\max}) \eta_1(\epsilon).
\end{align*}

We proceed by bounding the relative entropy between $Q_{\wstar}$ and $Q_{w}$ for all $w\neq \wstar$.  Let $\Qbar = P_{Y_r|X_{r}, Y^{r-1}}$ which is the effective channel (averaged over possible states) from the transmitter's perspective at time $r$. Also let $\Xstar  = e(\Wstar, y^{r-1}).$ Then we have the following lemma.
\begin{lem}\label{lem:Kl Q_w and Q bar}
Given an AVC, let $Q_w$ be as in \eqref{eq:Q_w}. Then, if $H_{r-1}< \epsilon$, the following inequality holds
\begin{align}\label{eq:remove w in D}
\max_w D_{KL}\Big( Q_{\wstar}~\|~ Q_{w} \Big) & \leq \max_{x} D_{KL}\Big(\Qbar(\cdot | \xstar, y^{r-1})~ \|~ \Qbar(\cdot| x, y^{r-1})\Big)+O(\eta_1(\epsilon)),
 \end{align} 
 where $\Qbar = P_{Y_r|X_{r}, Y^{r-1}}$ is the average channel at time $r$, and $\eta_1(\epsilon) = h_b^{-1}(\epsilon)$.
\end{lem}
With this lemma, we get the desired bound on the logarithmic drift of the entropy  
\begin{align*}
&\EE[\log H_{r-1}-\log H_{r}|\mathcal{F}_{r-1}]\leq \max_{x} D_{KL}\Big(\Qbar(\cdot | \xstar, y^{r-1})~ \|~ \Qbar(\cdot| x, y^{r-1})\Big)  + O(\eta_1(\epsilon)).
\end{align*}
 Therefore, the main lemma is proved. It remains to prove \eqref{eq:remove w in D}. 
\end{IEEEproof}
\begin{IEEEproof}[Proof of Lemma \ref{lem:Kl Q_w and Q bar}]
From the convexity of the relative entropy, the left-hand side term in \eqref{eq:remove w in D} equals to 
\begin{align}\label{eq:log drift bound 1}
\max_w D_{KL}\Big( Q_{\wstar}~\|~ Q_{w} \Big) = \sup_{\{\mu(w)\}_{w\neq \wstar}}D_{KL}\Big( Q_{\wstar}~\|~ \sum_{w\neq \wstar}\frac{\mu(w)}{1-\mu(\wstar)} Q_{w} \Big),
\end{align}
where the supremum is taken over all distributions $\mu(\cdot)$ on all $w\neq \wstar$ satisfying $\mu(w)\geq 0$ and $\sum_{w\neq \wstar} \mu(w) = 1-\mu(\wstar)$.
We upper bound the right-hand side of \eqref{eq:log drift bound 1} by approximating $Q_{\wstar}(y_r)$ and bounding $\sum_{w\neq \wstar}\frac{\mu(w)}{1-\mu(\wstar)} Q_{w}$ from below.


We start with lower-bounding the second term. For any $x\in\mathcal{X}$, define $\nu(x) \deq \prob{X_{r}=x | y^{r-1}}$.
Then, for any choice of $\mu(w), w\neq \wstar$, we have that
 \begin{align*}
\sum_{w \notin \Wstar }& \frac{\mu(w)}{1-\mu(\Wstar)} Q_{w}(y_r) =  \frac{1}{1-\mu(\Wstar)}\sum_{w \notin \Wstar} \prob{W= w, Y_r = y_r \big|~Y^{r-1}= y^{r-1} }\\
& \stackrel{(a)}{\geq} \frac{1}{1-\mu(\Wstar)}\sum_{\substack{w \notin  \wstar,\\ x\notin  \xstar}} \sum_{s\in\mathcal{S}} \prob{W= w, X_{r} = x, S_{r} = s, Y_r = y_r \big|~Y^{r-1}= y^{r-1} }\\
& \stackrel{(b)}{=} \frac{1}{1-\mu(\Wstar)}\sum_{\substack{w\\ x \notin \xstar}} \sum_{s\in\mathcal{S}} \prob{W= w, X_{r} = x, S_r = s, Y_r = y_r \big|~Y^{r-1}= y^{r-1} }\\
& =\frac{1}{1-\mu(\Wstar)}\sum_{\substack{x \notin  \xstar}} \sum_{s\in\mathcal{S}} \prob{X_{r} = x, S_r = s, Y_r = y_r \big|~Y^{r-1}= y^{r-1} }\\
& = \frac{1}{1-\mu(\Wstar)}\sum_{\substack{x\notin \xstar}} \sum_{s\in\mathcal{S}} \prob{X_{r} = x, S_r = s\big|~Y^{r-1}= y^{r-1} }Q(y_r | x,s)\\
& \stackrel{(c)}{=} \frac{1}{1-\mu(\Wstar)}\sum_{\substack{x\notin  \xstar}} \prob{X_{r} = x \big|~Y^{r-1}= y^{r-1} } \sum_{s\in\mathcal{S}} \prob{S_r = s\big|~Y^{r-1}= y^{r-1},X_{r} = x } Q(y_r | x,s)\\
&\stackrel{(d)}{=} \frac{1}{1-\mu(\Wstar)}\sum_{\substack{x\notin  \xstar}} \nu(x) \Qbar(y_r|x, y^{r-1})\\\numberthis \label{eq:Q1 Qbar bound 2}
& \stackrel{(e)}{\approx} \frac{1}{1-\nu(\xstar)}\sum_{\substack{x\notin  \xstar}} \nu(x) \Qbar(y_r|x, y^{r-1}),
 \end{align*}
 where (a) holds by summing over $x$ and $s$. Note that this is an inequality because the summation is not over all values of $x$. Equality (b) holds by removing the condition $w\neq \wstar$. This is an equality, because the probability of the event $\{W =\wstar, X_{r} \neq \xstar\}$ is zero. Equality (c) follows by breaking the joint probability on $(X_{r}, S_r)$ and further moving the summation on $s$. Equality (d) is due to the definition of $\nu$ and $\Qbar$. Lastly, (e) is due to the fact that $\nu(\xstar) \approx \mu(\wstar) \approx 1-\eta_1(\epsilon)$.

 Next, we approximate $Q_{\wstar}(y_r)$ by deriving a lower bound and an upper bound that are converging to each other. We start with the lower bound on $Q_{\wstar}(y_{r})$. Since $X_{r}=e(W, y^{r-1}),$  we have that 
\begin{align*}
Q_{\wstar}(y_{r}) &= \prob{Y_r = y_r \big|~ W =  \wstar, Y^{r-1}=y^{r-1}, X_{r} =  \xstar}\\
&= \sum_{s \in\mathcal{S}} \prob{Y_r = y_r, S_r = s \big|~ W = \wstar, Y^{r-1}=y^{r-1}, X_{r} = \xstar}\\\numberthis \label{eq:Q1 Qbar eq 1}
&= \sum_{s \in\mathcal{S}} \prob{S_r = s \big|~ W = \wstar, Y^{r-1}=y^{r-1}, X_{r} = \xstar }Q(y_r| s, \xstar)
\end{align*}
where the last equality holds because of the channel's probability rules.  Next, we bound the conditional probability on $S_r$ in the above summation. This quantity equals to 
 \begin{align}\label{eq:Q1 Qbar eq 2}
 \prob{S_r = s \big|~  Y^{r-1}=y^{r-1}, X_{r} = \xstar }\frac{\PP\{W= \wstar | Y^{r-1}=y^{r-1}, X_{r} = \xstar, S_r = s\}}{\PP\{W= \wstar | Y^{r-1}=y^{r-1}, X_{r} = \xstar\}}
 \end{align}
 The denominator is greater than $(1-\eta_1(\epsilon))$ because of the following argument: 
 \begin{align*}
\prob{W= \wstar | Y^{r-1}=y^{r-1}, X_{r} = \xstar} &= \frac{\PP\{W= \wstar, X_{r} = \xstar | Y^{r-1}=y^{r-1}\}}{\PP\{X_{r} = \xstar | Y^{r-1}=y^{r-1}\}}\\
 &= \frac{\PP\{W= \wstar| Y^{r-1}=y^{r-1}\}}{\PP\{X_{r} = \xstar | Y^{r-1}=y^{r-1}\}}\\
 &= \frac{\mu(\wstar)}{\PP\{X_{r} = \xstar | Y^{r-1}=y^{r-1}\}}\\
 &\geq 1-\eta_1(\epsilon),
 \end{align*}
 where the last inequality holds as the denominator is less than one and that $\mu(\wstar)\geq 1-\eta_1(\epsilon).$
 
As the nominator in \eqref{eq:Q1 Qbar eq 2} is less than one, then we get that 
\begin{align*}
 \eqref{eq:Q1 Qbar eq 2}\leq \frac{1}{1-\eta_1(\epsilon)}\prob{S_r = s \big|~  Y^{r-1}=y^{r-1}, X_{r} = \xstar }
 \end{align*} 
 Hence, from  \eqref{eq:Q1 Qbar eq 1}, we have that 
  \begin{align*}
Q_{\wstar}(y_{r}) & \leq \frac{1}{1-\eta_1(\epsilon)}\sum_{s \in \mathcal{S}} \prob{S_r = s \big|~  Y^{r-1}=y^{r-1}, X_{r} = \xstar } Q(y_r| s, \xstar)\\\numberthis \label{eq:Q1 Qbar bound 1}
& = \frac{1}{1-\eta_1(\epsilon)} \Qbar(y_r| \xstar, y^{r-1}).
 \end{align*}
 
With the bounds in \eqref{eq:Q1 Qbar bound 1} and \eqref{eq:Q1 Qbar bound 2}, the right-hand side of \eqref{eq:log drift bound 1} is bounded as 
 \begin{align*}
 D_{KL}\Big( Q_{\wstar}~\|~ \sum_{w\neq \wstar}\frac{\mu(w)}{1-\mu(\wstar)} Q_{w}& \Big)  =  \sum_{y_r}Q_{\Wstar}(y_{r}) \log \frac{Q_{\Wstar}(y_{r})}{\sum_{w\notin \Wstar} \frac{\mu(w)}{1-\mu(\Wstar)} Q_{w}(y_{r})}\\
&\stackrel{(a)}{\leq}  \sum_{y_r}Q_{\Wstar}(y_{r}) \log \frac{\frac{1}{1-\eta_1(\epsilon)}\Qbar(y_{r}|\xstar, y^{r-1})}{\sum_{x\notin \xstar} \frac{\nu_1(x)}{1-\nu_1(\xstar)} \Qbar(y_{r}|x, y^{r-1})}\\
&\stackrel{(b)}{\leq} \sum_{y_r}\Qbar(y_{r}|\xstar, y^{r-1}) \log \frac{\Qbar(y_{r}|\xstar, y^{r-1})}{\sum_{x\notin \xstar} \frac{\nu_1(x)}{1-\nu_1(\xstar)} \Qbar(y_{r}|x, y^{r-1})}-\log (1-\eta_1(\epsilon))\\\numberthis\label{eq:log drift bound 2}
& = D_{KL}\Big(\Qbar(\cdot | \xstar, y^{r-1})|| \sum_{x\neq \xstar} \frac{\nu_1(x)}{1-\nu_1(\xstar)} \Qbar(\cdot| x, y^{r-1})\Big)+O(\eta_1(\epsilon)),
 \end{align*}
 where (a) follows from \eqref{eq:Q1 Qbar bound 1} and \eqref{eq:Q1 Qbar bound 2}, and (b) follows from the following argument for bounding $Q_{\wstar}(y_{r})$ from below: 
\begin{align*}
Q_{\wstar}(y_{r}) &= \prob{Y_r = y_r \big|~ W = \wstar, Y^{r-1}=y^{r-1}, X_{r}= \xstar}\\
& \geq \prob{Y_r = y_r, W = \wstar \big|~ Y^{r-1}=y^{r-1}, X_{r} = \xstar}\\
& = \prob{Y_r = y_r \big|~ Y^{r-1}=y^{r-1}, X_{r} = \xstar} \prob{ W = \wstar \big|~ Y^{r-1}=y^{r-1}, X_{r} = \xstar, Y_r = y_r}\\
&=\Qbar(y_r|\xstar, y^{r-1}) \frac{\prob{ W = \wstar, X_{r} = \xstar,\big|~ Y^{r-1}=y^{r-1}, Y_r = y_r}}{\prob{X_{r} = \xstar\big|~ Y^{r-1}=y^{r-1}, Y_r = y_r}}\\
&\stackrel{(a)}{=}\Qbar(y_r|\xstar, y^{r-1}) \frac{\prob{ W = \wstar \big|~ Y^{r-1}=y^{r-1}, Y_r = y_r}}{\prob{X_{r}= \xstar\big|~ Y^{r-1}=y^{r-1}, Y_r = y_r}}\\
&\stackrel{(b)}{\geq } \Qbar(y_r|\xstar, y^{r-1})\prob{ W = \wstar \big|~ Y^{r-1}=y^{r-1}, Y_r = y_r}\\
& \stackrel{(c)}{=} \Qbar(y_r|\xstar, y^{r-1}) \mu(\wstar, y_r),\\\numberthis\label{eq:Q1 Qbar bound 3}
& \stackrel{(d)}{\geq} \Qbar(y_r|\xstar, y^{r-1})(1-\eta_2(\epsilon))
%
 \end{align*} 
where  (a) holds as $X_{r}$ is a function of $(W, Y^{r-1})$, (b) holds as the denominator is less than one, (c) is due to the definition of $\mu(w, y_r)$, and (d) holds from \eqref{eq:1-mu1 approximate} and by defining 
\begin{align*}
\eta_2(\epsilon) =\eta_1(\epsilon)\frac{\sum_{w\notin \Wstar} \frac{\mu(w)}{1-\mu(\Wstar)}   Q_{w}(y_{r})}{P(y_r|y^{r-1})}.
\end{align*}
 Lastly, by taking the supermum of \eqref{eq:log drift bound 2} over $\nu(x), x\neq \xstar$, we obtain the desired bound in \eqref{eq:remove w in D}. With that the proof is complete.
 
\end{IEEEproof}

\section{Proof of Lemma \ref{lem:bounding S}}\label{app:proof:lem:bounding S}
From the definition of $\{S_t\}_{t>0},$ we have that
\begin{align*}
S_t = \sum_{r=1}^{t\wedge \Tlepstldi} \frac{\J_r}{I} + \sum_{r=t\wedge \Tlepstldi+1}^{t\wedge \Tuepstldi} \frac{\log M}{I} \11\{H_{r-1} \geq \addvb{\sqrt{\epsilon}}\}+\sum_{r=t\wedge \Tuepstld+1}^{t} \frac{\D_r}{D} + \addvb{\sqrt{\epsilon}} \frac{N}{I}\11\{t\geq \Tuepstld\}
\end{align*}
Therefore, as $(T\vee \Tuepstldi) \geq  \Tuepstldi \geq  \Tlepstldi$, then
\begin{align*}
\EE\big[S_{T\vee \Tuepstldi}\big] &= \EE\bigg[ \sum_{r=1}^{\Tlepstldi} \frac{\J_r}{I} + \sum_{r=\Tlepstldi+1}^{\Tuepstldi} \frac{\log M}{I} \11\{H_{r-1} \geq \addvb{\sqrt{\epsilon}}\}+\sum_{r=\Tuepstld+1}^{T\vee \Tuepstldi} \frac{\D_r}{D}\bigg] + \addvb{\sqrt{\epsilon}} \frac{N}{I}.
\end{align*}
As for the first summation, after multiplying and dividing by $\EE[\Tlepstldi]$, we have that
\begin{align*}
\EE\Big[\sum_{r=1}^{\Tlepstldi} \frac{\J_r}{I}\Big] = \frac{\EE[\Tlepstldi]}{I}  \bigg(\frac{1}{\EE[\Tlepstldi]} \EE\Big[\sum_{r=1}^{\Tlepstldi} \J_r\Big] \bigg) = \EE[\Tlepstldi],
\end{align*}
where the last equality follows by setting $I$ as in the statement of the lemma. Similarly, the third summation is bounded as in the following 
\begin{align*}
\EE\Big[\sum_{r=\Tuepstld+1}^{T\vee \Tuepstldi} \frac{\D_r}{D}\Big] = \frac{\EE[T-\Tlepstldi]}{D}  \bigg(\frac{1}{\EE[T-\Tlepstldi]}\EE\Big[\sum_{r=\Tuepstld+1}^{T\vee \Tuepstldi} \D_r\Big] \bigg) = \EE[T-\Tlepstldi],
\end{align*}
where the first equality holds after multiplying and dividing by $\EE[T-\Tlepstldi]$, and the second equality follows by setting $D$ as in the statement of the lemma. As a result,
\begin{align*}
\EE\big[S_{T\vee \Tuepstldi}\big] & \leq \EE[T] + \frac{\log M}{I}\EE\bigg[ \sum_{r=\Tlepstldi+1}^{\Tuepstldi}  \11\{H_{r-1} \geq \addvb{\sqrt{\epsilon}}\}\bigg] + \addvb{\sqrt{\epsilon}} \frac{N}{I}\\\numberthis \label{eq:bounding S eq 1}
&{\leq} \EE[T] + \frac{\log M}{I}\EE\bigg[ \sum_{r=\Tlepstldi+1}^{N}  \11\{H_{r-1} \geq \addvb{\sqrt{\epsilon}}\}\bigg] + \addvb{\sqrt{\epsilon}} \frac{N}{I}
\end{align*}
where the inequality follows as $\Tuepstldi \leq N$.  Next, we bound the remaining summation. By iterative expectation we have that
\begin{align*}
\EE\Big[\sum_{r=\addvb{\Tlepstldi}+1}^{N} \11\{H_{r-1} \geq \addvb{\sqrt{\epsilon}}\}\Big] &= \EE_{\addvb{\Tlepstldi}}\Big[\sum_{r=\addvb{\Tlepstldi}+1}^{N} \EE\big[\11\{H_{r-1} \geq \addvb{\sqrt{\epsilon}}\} ~\big| \addvb{\Tlepstldi} \big]\Big]\\
& = \EE_{\addvb{\Tlepstldi}}\Big[\sum_{r=\addvb{\Tlepstldi}+1}^{N} \PP\big(H_{r-1} \geq \addvb{\sqrt{\epsilon}}  ~\big| \addvb{\Tlepstldi} \big)\Big]\\
&\stackrel{(a)}{\leq}  \EE_{\addvb{\Tlepstldi}}\Big[\sum_{r=\addvb{\Tlepstldi}+1}^{N} \PP\Big( \sup_{\addvb{\Tlepstldi} \leq t \leq N-1 }H_{t} \geq \addvb{\sqrt{\epsilon}} ~\big| \addvb{\Tlepstldi} \Big)\Big]\\
&=  \EE_{\addvb{\Tlepstldi}}\Big[\PP\Big( \sup_{ \addvb{\Tlepstldi} \leq t \leq N-1 }H_{t} \geq \addvb{\sqrt{\epsilon}} ~\big|\addvb{\Tlepstldi}\Big) \sum_{r=\addvb{\Tlepstldi}+1}^{N} \11 \Big]\\
&\stackrel{(b)}{\leq}  \EE_{\addvb{\Tlepstldi}}\Big[N  \PP\Big( \sup_{ \addvb{\Tlepstldi} \leq t \leq N-1 }H_{t} \geq \addvb{\sqrt{\epsilon}}~\big| \addvb{\Tlepstldi} \Big)\Big]\\\numberthis \label{eq:bounding S eq 2}
&\stackrel{(c)}{=}  N \PP\Big( \sup_{ \addvb{\Tlepstldi} \leq t \leq N-1 }H_{t} \geq \addvb{\sqrt{\epsilon}} \Big),
\end{align*}
where (a) follows from taking the supremum over all  $H_{r-1}$ appearing in the summation. Inequality (b) follows as the summation is less than $N-\addvb{\Tlepstldi} $ which is smaller than $N$. Lastly, (c) holds by taking the expectation of the conditional probability. We proceed with the following lemma which is a variant of Doob's maximal inequality for super-martingales. 
\begin{lem}[Maximal Inequality for Supermartingales]\label{lem:super martingale}
Let $\{M_t\}_{t>0}$ be a non-negative supermartingale w.r.t a filtration $\{\mathcal{F}_t\}_{t>0}$ . If $\tau$ is a bounded stopping time w.r.t this filtration, then  the following inequality holds for any constant $c>0$
\begin{align*}
\prob{\sup_{t\geq \tau} M_t >c } \leq \frac{\EE[M_\tau]}{c}
\end{align*}
\end{lem}
\begin{IEEEproof}
Define $S\deq \inf\set{t>0: t\geq \tau, M_t >c}$. Note that $S$ is a stopping time. Since $\{M_t\}_{t>0}$ is non-negative, then for any fixed $n\in \NN$, we have that 
\begin{align*}
M_{n \wedge S} \geq c \11\Big\{\sup_{\tau \leq t \leq n} M_t >c\Big\}. 
\end{align*}
Therefore, taking the expectation of both sides and rearranging the terms gives the following inequality 
\begin{align}\label{eq:lem super martingale 1}
\PP\Big\{\sup_{\tau \leq t \leq n} M_t >c\Big\} \leq \frac{\EE[M_{n \wedge S}]}{c}.
\end{align}
Since $\{M_t\}_{t>0}$ is a super-martingale and that $\tau \leq S$, then $\EE[M_{n \wedge \tau}] \geq \EE[M_{n \wedge S}].$ Therefore, we can write 
\begin{align*}
\PP\Big\{\sup_{\tau \leq t \leq n} M_t >c\Big\} \leq \frac{\EE[M_{\tau}]}{c}.
 \end{align*} 
This is because if $n<\tau$ then the left-hand side is zero and the inequality holds trivially. When $n \geq \tau$, using the above argument, the right-hand side of \eqref{eq:lem super martingale 1} is less than $\frac{\EE[M_{n \wedge \tau}]}{c} = \frac{\EE[M_{\tau}]}{c}$.

Next, taking the limit $n\rightarrow \infty$ and from monotone convergence theorem we get that 
\begin{align*}
\PP\Big\{\sup_{\tau \leq t } M_t >c\Big\} &= \PP\Big\{\medcup_{n>0} \big\{\sup_{\tau \leq t \leq n} M_t >c\big\} \Big\}\\
&= \lim_{n\rightarrow \infty} \PP\Big\{\sup_{\tau \leq t \leq n} M_t >c\Big\}\\
&\leq \frac{\EE[M_{\tau}]}{c},
\end{align*}
where the second equality follows from the continuity of the probability measure. 
With that the proof is complete. 
\end{IEEEproof}
Note that $\{H_t\}_{t>0}$ is a super martingale. Therefore, from Lemma \ref{lem:super martingale}, we have that 
\begin{align*}
\PP\Big( \sup_{ \addvb{\Tlepstldi}  \leq t \leq N-1 }H_{t} \geq \addvb{\sqrt{\epsilon}} \Big) \leq \frac{\EE[H_{\addvb{\Tlepstldi} }]}{\addvb{\sqrt{\epsilon}}}
\end{align*}
If $\addvb{\Tlepstldi} <N$, then by definition of this stopping time  $H_{\addvb{\Tlepstldi}} \leq \epsilon$; otherwise $\addvb{\Tlepstldi} = N$ which implies that $H_{\addvb{\Tlepstldi}} = H_{N}$. However, as $T\leq N$, then 
\begin{align*}
\EE[H_{N}]\leq \EE[H_T] \leq h_b(P_e) + P_e\log M_1M_2 \leq \epsilon,
\end{align*}
where the second inequality follows from Fano's and the last inequality holds as $P_e \ll \epsilon$. Consequently,
\begin{align*}
\PP\Big( \sup_{ \addvb{\Tlepstldi} \leq t \leq N-1 }H_{t} \geq \addvb{\sqrt{\epsilon}} \Big) \leq \frac{\epsilon}{\addvb{\sqrt{\epsilon}}}= \addvb{\sqrt{\epsilon}}.
\end{align*}
Therefore, using this inequality in \eqref{eq:bounding S eq 2} we obtain that 
\begin{align*}
\EE\Big[\sum_{r=\addvb{\Tlepstldi}+1}^{N} \11\{H_{r-1} \geq \addvb{\sqrt{\epsilon}}\}\Big] \leq N \addvb{\sqrt{\epsilon}}.
\end{align*}
Thus, from \eqref{eq:bounding S eq 1}, we obtain that 
\begin{align*}
\EE[S_{T\vee \Tuepstldi}]  & \leq \EE[T] + \frac{\log M}{I} \big(\addvb{\sqrt{\epsilon} N} \big)+ \addvb{\sqrt{\epsilon}}\frac{N}{I}.
\end{align*}
Hence, factoring $\EE[T]$ gives the following inequality 
\begin{align*}
\EE[S_{T\vee \Tuepstldi}]  & \leq \EE[T]( 1 + V(\epsilon, N)),
\end{align*}
where $V(\epsilon,N) =\frac{R_i}{I} \big( \addvb{\sqrt{\epsilon}} N \big)+ \addvb{\sqrt{\epsilon}} \frac{N}{\EE[T] I}.$ 
 Hence, the proof is complete. 
\end{document}